\documentclass[nonacm, sigconf]{acmart} %

\usepackage{tikz}
\usepackage{amsmath}

\usepackage{filecontents}

\usepackage{subcaption}

\usepackage{booktabs}
\usepackage{paralist}
\usepackage[group-separator={,}, per-mode=symbol]{siunitx}
\usepackage{xspace}
\usepackage{eurosym}
\usepackage{makecell}

\usepackage{xurl}

\usepackage{graphicx}
\usepackage{color, colortbl}
\usepackage[ruled,vlined]{algorithm2e}
\SetKwComment{Comment}{$\triangleright$\ }{}
\usepackage{mathtools}

\usepackage{amsfonts}
\usepackage{amsmath}
\usepackage{bm}
\usepackage{threeparttable, tablefootnote}
\usepackage{multirow}
\usepackage{lipsum}                     %

\usepackage{soul}
\usepackage{comment}
\usepackage{dirtytalk}
\usepackage[nolist]{acronym}
\usepackage{csquotes}
\usepackage{enumitem}
\usepackage{hyperref}
\usepackage{float}

\usepackage{circledtext}
\circledtextset{resize=real}

\AtBeginDocument{%
  }

\copyrightyear{2026}
\acmYear{2026}
\acmConference[CCS '26] {Proceedings of the 2026 ACM SIGSAC Conference on Computer and Communications Security}{November 15--19, 2026}{The Hague, Netherlands.}
\acmBooktitle{Proceedings of the 2026 ACM SIGSAC Conference on Computer and Communications Security (CCS '26), November 15--19, 2026, The Hague, Netherlands}
\acmISBN{979-8-4007-2871-6/2026/11} 
\acmDOI{10.1145/3830454.3846807} %

\newcommand{\cf}{\textit{cf}.\xspace}
\newcommand{\etal}[0]{\textit{et~al.}\xspace}
\newcommand{\eg}{\textit{e.g.}\xspace}
\newcommand{\ie}{\textit{i.e.}\xspace}

\newcommand{\tablecircle}[1]{\circledtext*[height=2.3ex,charshrink=0.7]{#1}}
\newcommand{\tablecirclesmall}[1]{\circledtext*[height=1.9ex,charshrink=0.7]{#1}}

\newcommand{\Paragraph}[1]{\smallskip\noindent{\bf #1.}}

\begin{document}

\title{Talking to the Airgap: Exploiting Radio-Less Embedded Devices as Radio Receivers}

\author{Paul Staat$^{*}$}
 \affiliation{%
   \institution{Ruhr University Bochum}
   \city{Bochum}
   \country{Germany}}
 \email{paul.staat@rub.de}

\author{Daniel Davidovich$^{*}$}
\affiliation{%
   \institution{Max Planck Institute for Security and Privacy}
   \city{Bochum}
   \country{Germany}}
\email{daniel.davidovich@mpi-sp.org}

\author{Christof Paar}
 \affiliation{%
   \institution{Max Planck Institute for Security and Privacy}
   \city{Bochum}
   \country{Germany}}
\email{christof.paar@mpi-sp.org}

\renewcommand{\shortauthors}{Paul Staat, Daniel Davidovich, \& Christof Paar}

\begin{abstract}
Extended paper version, including three additional appendices.
\vspace{-0.2cm}\\

Physical isolation from external networks -- an \textit{airgap} -- aims to minimize exposure to remote attacks. Yet capable adversaries still achieve code execution on air-gapped systems, and prior work has shown that they can then wirelessly \textit{exfiltrate} data via unintended emissions. In this work, we demonstrate the reverse direction: malicious code on an embedded device enables wireless \textit{infiltration} of air-gapped systems, granting attackers command-and-control over compromised targets. Leveraging physical effects previously studied in the context of \ac{EMI}, we show that parasitic \ac{RF} sensitivity in \ac{PCB} traces and on-chip \acp{ADC} turns commodity embedded devices into inadvertent radio receivers. Unlike prior infiltration techniques, our approach requires no dedicated sensors (\eg, microphones, LEDs, or temperature sensors) and works in non-line-of-sight scenarios.
In our evaluation, an ordinary microcontroller evaluation board reliably recovers communication signals from tens of meters at data rates of up to \SI{100}{kbps}. Applying a systematic methodology to discover such device-intrinsic \ac{RF} sensitivity, we evaluate twelve commercial embedded devices and two custom prototypes, finding that \emph{all} exhibit reception capabilities in the 300--1000~MHz range. Our findings challenge the assumption that embedded devices without radios lack an inbound radio paths and call for air-gap threat models that account for both emission-based leakage and unintended reception.
\end{abstract}

%
%\begin{CCSXML}
%<ccs2012>
%   <concept>
%       <concept_id>10010583.10010588.10011669</concept_id>
%       <concept_desc>Hardware~Wireless devices</concept_desc>
%       <concept_significance>300</concept_significance>
%       </concept>
%   <concept>
%       <concept_id>10002978.10003001.10003003</concept_id>
%       <concept_desc>Security and privacy~Embedded systems security</concept_desc>
%       <concept_significance>500</concept_significance>
%       </concept>
% </ccs2012>
%\end{CCSXML}
%
%\ccsdesc[300]{Hardware~Wireless devices}
%\ccsdesc[500]{Security and privacy~Embedded systems security}
%
%
%\keywords{Embedded Security, Intentional Electromagnetic Interference, Wireless Security%
%} %

\maketitle
\def\thefootnote{*}\footnotetext{These authors contributed equally to this work.}

\begin{acronym}[JSONP]
\setlength{\itemsep}{0.2em}
\acro{ASK}{amplitude-shift keying}
\acro{SDR}{software-defined radio}
\acro{EMI}{electromagnetic interference}
\acro{EMC}{electromagnetic compatibility}
\acro{IEMI}{intentional electromagnetic interference}
\acro{RFI}{radio-frequency interference}
\acro{VDU}{visual display unit}
\acro{CRT}{cathode-ray tube}
\acro{AM}{amplitude-modulated}
\acro{FM}{frequency-modulated}
\acro{OOK}{on-off keying}
\acro{FSK}{frequency shift keying}
\acro{RF-PWM}{radio-frequency pulse-width modulation}
\acro{DRAM}{dynamic random-access memory}
\acro{ES}{embedded system}
\acro{PC}{personal computer}
\acro{TV}{television}
\acro{MPU}{microprocessor unit}
\acro{MCU}{microcontroller unit}
\acro{ASIC}{application-specific integrated circuit}
\acro{FPGA}{field-programmable gate array}
\acro{CPU}{central processing unit}
\acro{HW}{hardware}
\acro{SW}{software}
\acro{TRF}{tuned radio frequency}
\acro{IF}{intermediate frequency}
\acro{RF}{radio frequency}
\acro{LO}{local oscillator}
\acro{ADC}{analog-to-digital converter}
\acro{DAC}{digital-to-analog converter}
\acro{IC}{integrated circuit}
\acro{SPI}{serial peripheral interface}
\acro{TI}{Texas Instruments}
\acro{EM}{electromagnetic}
\acro{IR}{infrared}
\acro{ESD}{electro-static discharge}
\acro{AC}{alternating current}
\acro{DC}{direct current}
\acro{IMU}{inertial measurement unit}
\acro{UAV}{unmanned aerial vehicle}
\acro{ESC}{electronic stability control}
\acro{NTC}{negative temperature coefficient}
\acro{PTC}{positive temperature coefficient}
\acro{TEM}{transverse electromagnetic}
\acro{I2C}{Inter-Integrated Circuit}
\acro{ACK}{acknowledgement}
\acro{ABS}{antilock braking system}
\acro{DPI}{direct power injection}
\acro{BJT}{bipolar junction transistor}
\acro{JFET}{junction-gate field-effect transistor}
\acro{CMOS}{complementary metal–oxide–semiconductor}
\acro{UART}{universal asynchronous receiver-transmitter}
\acro{USB}{Universal Serial Bus}
\acro{HDMI}{High-Definition Multimedia Interface}
\acro{CAN}{Controller Area Network}
\acro{PWM}{pulse-width modulation}
\acro{GPIO}{general-purpose input/output}
\acro{RC}{remote controlled}
\acro{DMA}{direct memory access}
\acro{GUI}{graphical user interface}
\acro{SCPI}{Standard Commands for Programmable Instruments}
\acro{API}{application programming interface}
\acro{VSWR}{voltage standing wave ratio}
\acro{ISR}{interrupt service routine}
\acro{LSB}{least significant bit}
\acro{LOS}{line of sight}
\acro{CW}{continuous wave}
\acro{RMS}{root mean square}
\acro{DUT}{device under test}
\acro{SMA}{SubMiniature version A}
\acro{SNR}{signal-to-noise ratio}
\acro{PCB}{printed circuit board}
\acro{OEM}{original equipment manufacturer}
\acro{OS}{operating system}
\acro{FSPL}{free-space path loss}
\acro{MLS}{maximum length sequence}
\acro{BER}{bit error rate}
\acro{SCR}{silicon controlled rectifier}
\acro{ISM}{industrial, scientific and medical}
\acro{SCADA}{supervisory control and data acquisition}
\acro{IoT}{internet of things}
\acro{LED}{light-emitting diode}
\acro{TEE}{trusted execution environment}
\acro{UHF}{ultra-high frequency}
\acro{LNA}{low-noise amplifier}
\acro{PLL}{phase-locked loop}
\acro{VCO}{voltage-controlled oscillator}
\acro{UWB}{ultra-wideband}
\acro{NLOS}{non line-of-sight}
\end{acronym}

\section{Introduction}

Computing systems in security-critical roles -- such as cryptocurrency wallets and industrial control systems -- are often isolated from external networks behind an \emph{airgap}~\cite{guriBridgewareAirgapMalware2018} to block remote attacks entirely.

Airgaps raise the cost of remote attack but do not preclude it: capable adversaries -- including nation-state actors -- still penetrate air-gapped environments through hardware Trojans~\cite{puschnerRedTeamVs2023}, supply-chain compromise~\cite{robertson_big_2018}, or portable media~\cite{kushnerRealStoryStuxnet2013}, as illustrated by the 2010 \emph{Stuxnet} attack on Iranian uranium enrichment centrifuges~\cite{kushnerRealStoryStuxnet2013} and the 2024 weaponization of Hezbollah pagers and radios~\cite{hezbollahPagers2024}. Once implanted, sophisticated malware is seldom fire-and-forget. An inbound channel enables the attacker to update payloads as defenses evolve, rotate cryptographic keys, retarget, coordinate effects with external events, or retire implants before forensic discovery --capabilities that Stuxnet exercised across multiple software revisions during its operational lifetime~\cite{kushnerRealStoryStuxnet2013}. Reusing the original insertion vector for such updates is often impractical: portable-media drops require renewed physical access, supply-chain insertions are typically one-shot, and any of these vectors becomes foreclosed once defenders are alerted. Attackers therefore benefit from alternative covert inbound channels. Among these, a radio channel is uniquely attractive: it operates at a distance without physical access, is not human perceptible, and -- unlike acoustic or optical covert channels -- propagates through walls and enclosures, allowing command-and-control from outside the secured perimeter.

As a concrete example, consider a malicious implant delivered once via supply-chain compromise into an air-gapped industrial controller, limited to pre-defined behavior, \eg, triggering a denial-of-service at a fixed future date. Changing this behavior requires communicating with the implant post-deployment: an \ac{RF} channel as described in this work lets the attacker transmit a signal from outside the secured perimeter, which the implant receives and decodes, \eg, to shift the trigger date.

\begin{figure}%
    \centering
    \includegraphics[width=0.9\linewidth]{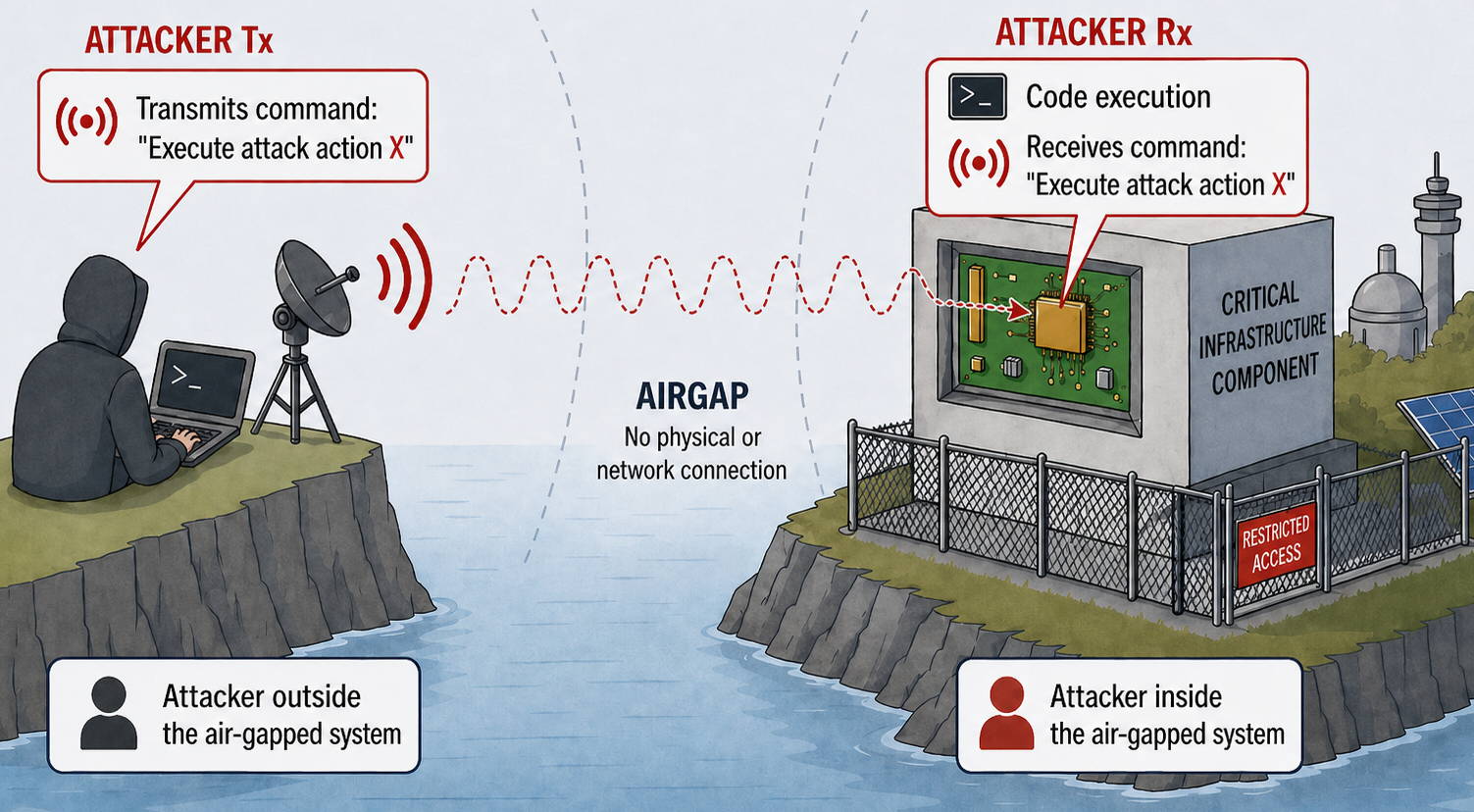}
    \caption{Attack scenario considered in this work. The attacker executes code on an air-gapped embedded device and receives radio commands.}
    \label{fig:threat_model}
    \Description{}
\end{figure}

The challenge of \textit{exfiltrating} data from air-gapped systems has been studied extensively for over 25~years, covering electromagnetic, magnetic, acoustic, vibrational, optical, and thermal leakage~\cite{kuhnSoftTempestHidden1998, guriBitWhisperCovertSignaling2015, kuhnapfelLaserSharkEstablishingFast2021, liSpiralSpyExploringStealthy2022, xuDiskSpyExploringLongRange2025, camuratiNoiseSDRArbitraryModulation2022}. \textit{Infiltration} has received far less attention. As summarized in \autoref{tab:related_work_infiltrate}, the few existing approaches rely on dedicated sensing hardware (\eg, temperature sensors, LEDs, microphones, or cameras), require close proximity or \ac{LOS}, and reach only low data rates.

\begin{table}[!htb]
\small
\centering
\caption{Communication Into Airgap Systems.}
\label{tab:related_work_infiltrate}
\begin{tabular}{@{}rcccc@{}}
\toprule
\textbf{Signal} & \textbf{\makecell{Sensor}} & \textbf{\makecell{Max.\\dist.}} & \textbf{\makecell{No\\LOS}} & \textbf{\makecell{Bitrate\\in bps}}\\
\midrule
Thermal \cite{guriBitWhisperCovertSignaling2015} & Temperature & 0.4 m & no & 0.002\\
Laser \cite{kuhnapfelLaserSharkEstablishingFast2021} & LED & 25~m & no  & 18,200\\
Ultrasonic \cite{guriMosquitoCovertUltrasonic2018} & Microphone & 8 m & no  & 166\\
Infrared \cite{guriAIRJumperCovertAirgap2019} & Camera & 130~m & no & 100\\
Radio~\cite{kasmiAirgapLimitationsBypass2016} & Temperature & 20~m & (yes) & 2.5\\
\midrule
\textbf{\makecell[r]{Radio\\(this work)}} & \makecell[c]{none\\(device itself)} & 20~m & yes & \makecell[c]{1,000\\(up to 100k)}\\
\bottomrule 
\end{tabular}
\end{table}

\paragraph{Threat model.} Our threat model -- attacker-controlled code execution on an embedded target -- matches that of the exfiltration literature~\cite{kuhnSoftTempestHidden1998, guriBitWhisperCovertSignaling2015, camuratiNoiseSDRArbitraryModulation2022}. How such access is obtained, be it through supply-chain compromise, hardware implants, boot ROM exploits, or malicious updates during a maintenance window, is orthogonal to our contribution. We stress that secure boot and firmware integrity mechanisms do not render this threat model theoretical: even protected systems can be breached through fault injection, voltage or clock glitching, and hardware implants. Stuxnet itself propagated signed, trusted-looking code onto hardened industrial controllers~\cite{kushnerRealStoryStuxnet2013}.

\paragraph{Our approach.}
We show that the physical structure of a compromised device alone -- without any sensor, transceiver, or component intended for wireless reception -- suffices to receive over-the-air communication signals. This sets us apart from prior covert channels relying on inherently signal-sensitive components such as cameras or microphones, and exposes an attack vector that current exfiltration-focused airgap threat models do not address. Restricted to code execution on the target, the attacker must build a software-defined radio receiver under extreme resource constraints, raising three core challenges:

\begin{enumerate}
    \item \label{chal:absence_wireless} \textbf{Absence of wireless components.} Without dedicated hardware, receiver building blocks -- antenna, amplifiers, and frequency conversion -- must be substituted.
    \item \label{chal:finding_sensitivity} \textbf{Finding reception sensitivities.} Identifying peripheral configurations with exploitable \ac{RF} sensitivity requires a systematic discovery methodology.
    \item \label{chal:lightweight_rx} \textbf{Constrained resources.} Reception must tolerate imperfect signals on tightly resource-constrained devices.
\end{enumerate}

For \ref{chal:absence_wireless}, we observe that ordinary \ac{PCB} traces unintentionally act as \ac{UHF} antennas (\SI{300}{}--\SI{1000}{MHz}), while device-intrinsic non-linearities at GPIO pins yield frequency down-conversion to baseband sampled by on-chip \acp{ADC}. For \ref{chal:finding_sensitivity}, we propose a methodology that sweeps peripheral configurations and observes the device's response with and without a controlled \ac{RF} signal, estimating \ac{SNR} to identify configurations that maximize sensitivity. Testing 14 real-world devices -- including commercial hardware wallets and a drone -- we find that \emph{all} exhibit significant radio reception, stable across time, location, orientation, and device samples. For \ref{chal:lightweight_rx}, we experimentally demonstrate that lightweight software-defined receiver algorithms recover transmitted data error-free at \SI{1}{kbps} -- enough for typical command-and-control payloads to be delivered within seconds of attacker presence.

Extending the rationale of \ac{IEMI}-based sensor manipulation~\cite{jiangPowerRadioManipulateSensor2025, lavauSecuringTemperatureMeasurements2023, tuTrickHeatManipulating2019} -- where Kasmi~\etal~\cite{kasmiAirgapLimitationsBypass2016} used kilowatt-level \ac{RF} to induce a \SI{2.5}{bps} channel via a temperature sensor -- we treat the compromised device \emph{itself} as an effective sensor. Unintended electromagnetic sensitivity in \ac{PCB} traces and on-chip circuitry, long known as a source of interference~\cite{kaurElectromagneticInterference2011}, thereby becomes an inbound communication channel: software on a compromised \ac{MCU} can receive signals arriving with as little as \SI{0.2}{\mW}, achieving \SI{1}{kbps} over \SI{20}{\m}, including in non-line-of-sight scenarios -- roughly $400\times$ the bitrate of~\cite{kasmiAirgapLimitationsBypass2016} at several orders of magnitude lower transmit power.

Our findings challenge conventional wisdom in hardware security: designers cannot rule out radio reception simply because no sensor is present, and threat assessments must account for offensive capabilities that go beyond the components a system was designed with. Beyond exposing the attack surface, our methodology offers a constructive path forward: by enabling the proactive identification of subtle \ac{RF} sensitivities, it supports the design of embedded systems with improved electromagnetic resilience that actively prevents unintended reception capabilities. We make the following contributions:
\begin{itemize}[nosep]
\item We demonstrate that attackers can repurpose air-gapped embedded devices into unintended radio receivers supporting inbound wireless command-and-control at kilobit-per-second rates.
\item We introduce an automated methodology for discovering unintended \ac{RF} reception paths, and apply it to 14 platforms (12 commercial devices -- including hardware wallets and a drone -- and two custom \acp{PCB}).
\item We systematically characterize reception behavior as a function of transmit power, carrier frequency, device orientation, timing, and hardware variation, quantifying achievable bitrates, bit-error rates, and link distances.
\item We evaluate signal-integrity-aware \ac{PCB} layout and shielding techniques as countermeasures.
\item We release firmware for sensitivity testing on eight STM32 \acp{MCU}, design files for our custom \acp{PCB}, and the software-based receiver implementation.
\end{itemize}
\section{Preliminaries}
\label{sec:background}

In this section, we provide technical background on radio receivers and sensitivity of electronic components for electromagnetic signals and define the threat model. 

\subsection{Technical Background}

\begin{figure}
    \begin{subfigure}{0.38\textwidth}
        \centering
        \includegraphics[width=\linewidth]{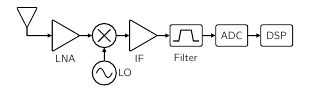}
        \caption{}
        \label{fig:receiver:ideal}
    \end{subfigure}\\
    \begin{subfigure}{0.38\textwidth}
        \centering
        \includegraphics[width=\linewidth]{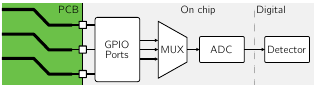}
        \caption{}
        \label{fig:receiver:mcu}
    \end{subfigure}
    \caption{(a)~Classical frequency-tuned \ac{RF} receiver. (b)~Embedded device-based radio receiver.}
    \label{fig:ideal_receiver}
\end{figure}

\Paragraph{Conventional Radio Reception}
Conventional radio receivers, such as the direct conversion architecture shown in~\autoref{fig:receiver:ideal}, consist of several fundamental components. An \emph{antenna} captures radio signals from the air and converts them into electrical signals which are then amplified by a \emph{\ac{LNA}} to improve sensitivity. A \emph{mixer}, driven by a \emph{local oscillator}, shifts the frequency of the received \ac{RF} signal to an intermediate frequency or baseband. A \emph{band-select filter} removes out-of-band signals before the signal is digitized by an \emph{\ac{ADC}}. Then, \emph{digital signal processing} enables software-defined reception~\cite{goldsmithWirelessCommunications2005}. Modern receiver designs can also perform frequency translation and filtering entirely in the digital domain through direct-RF sampling where the output of the \ac{LNA} is digitized directly.

\Paragraph{Electromagnetic Sensitivity of Electronics}

The electronic circuits of computer systems are inherently vulnerable to \ac{EMI} due to their physical nature. \ac{EMI} can originate from both natural and man-made sources, degrading the performance and reliability of an electronic system~\cite{estevesElectromagneticInterferenceInformation2023, gilCharacterizationModellingEMI2012}. Interference can couple into a device through conductive, inductive, capacitive, or radiative mechanisms~\cite{kaurElectromagneticInterference2011}. Wires such as \ac{PCB} traces can act as antennas, allowing electromagnetic energy to reach digital or analog components~\cite{archambeault2002pcb}. Although peripheral sampling rates are often much lower than the signal frequency of \ac{EMI}, nonlinearities in amplifiers, \ac{ADC}s, and other components can mix, rectify, or demodulate high-frequency disturbances, producing low-frequency variations or DC offsets that can propagate through and affect the system~\cite{jiangPowerRadioManipulateSensor2025,richelli2016emi}.
On digital data lines, this may produce bit flips~\cite{zhangElectromagneticSignalInjection2023}, on analog signal lines, sensor readings or \ac{ADC} outputs can be distorted~\cite{richelli2016emi, pahlIntendedElectromagneticInterference2021}.

This topic is a major concern in safety-critical domains, including medical, aviation, and automotive applications~\cite{kaurElectromagneticInterference2011}. Thus, \ac{EMC} has been studied extensively, with a large body of research~\cite{ramdani2009electromagnetic, redl1996power} and corresponding regulatory frameworks~\cite{codeOfFederalRegulationPart15, europeEMC, emcSociety}.

\subsection{Threat Model}

\Paragraph{Attacker Capabilities}
The adversary has software-only access to a processor of an air-gapped embedded device such as a \ac{MCU} or \ac{FPGA} and obtained low-level code execution privileges. This includes full control over software and peripheral configuration, as well as access to on-chip \ac{ADC} readings. However, the adversary cannot alter the device's hardware, add external components, or modify the physical environment of the target. In addition, the adversary can position an external radio transmitter in range of the target and can adjust its transmission parameters, including power, frequency, and modulation.

\Paragraph{Goals}
The adversary seeks to establish a covert wireless communication channel into the air-gapped system. Specifically, the objective is to transmit arbitrary digital data from the external transmitter to the compromised device using only existing hardware capabilities.

\Paragraph{Assumptions}
The attacker may select any transmitter position which allows sufficient \ac{RF} signal power to reach the target device, implying that the physical environment allows signal propagation to the target device, including non-\ac{LOS} scenarios. The attacker can transmit arbitrary waveforms and does not have to comply with regulatory constraints on signal strength or spectrum use. 

The compromised target device has neither radio nor sensor capabilities by design. Since the attacker has no physical access to the device, they cannot make hardware alterations. However, we assume the attacker is aware of the hardware design of the target device, \eg, being in possession of design files for custom hardware or identical commodity hardware.

\section{Embedded Devices as Radio Receivers}

In this work, we investigate how embedded devices can be exploited as unintended wireless receivers to realize command-and-control of compromised air-gapped systems. Our hypothesis is that embedded systems can exhibit unintentional \ac{EMI} sensitivity when exposed to \ac{RF} signals. A key challenge lies in identifying such sensitivities -- if they exist -- and determining whether they support the reception of communication waveforms. To address this, we present a systematic framework and experimental setup to detect \ac{RF} sensitivities in real-world devices, forming the foundation for their subsequent characterization~(\autoref{sec:sensitivity_experiments}) and practical exploitation for wireless communication~(\autoref{sec:communication}).

\subsection{Our Idea}
Motivated by the literature on \ac{EMI}~\cite{kaurElectromagneticInterference2011, gilCharacterizationModellingEMI2012, mathur2020electromagnetic}, \ac{IEMI}~\cite{selvarajElectromagneticInductionAttacks2018, pahlIntendedElectromagneticInterference2021, tuTrickHeatManipulating2019}, and fault injection~\cite{cuiBADFETDefeatingModern2017, dehbaouiElectromagneticTransientFaults2012}, we hypothesize that embedded devices even with minimal hardware capabilities can act as basic radio receivers. Unlike prior works, which consider electromagnetic interference primarily in terms of system disruption or destructive effects, we focus on \textit{detectable} \ac{RF} sensitivities -- subtle effects that leave device operation intact while reliably making incident \ac{RF} energy measurable. For this, we leverage simple \acp{ADC} found in virtually every modern \ac{MCU} as the basic atomic primitive to bridge between the physical and digital domains. However, an \ac{ADC} alone does not make an \ac{RF} receiver: whether and how it responds to \ac{RF} stimuli depends on coupling mechanisms in the board and chip. To discover useful sensitivities we perform an exhaustive search across the following three key dimensions.

\Paragraph{Reception Paths}
For a given \ac{DUT}, the available reception paths are defined as the set of all possible connections that any on-chip \ac{ADC} can be connected to. This may include \ac{GPIO} pins, internal sensors such as temperature and voltage monitors, connections to other peripherals such as operational amplifiers, and reserved (undocumented) configurations. The available reception paths of a \ac{DUT} are determined by the number of \acp{ADC} and the number of MUX connections, see~\autoref{fig:receiver:mcu}.

\Paragraph{Path Configurations}
For any reception path that eventually connects to a physical pin of the \ac{MCU}, the \ac{GPIO} front end provides additional configuration options that may affect the electrical characteristics, such as pin direction, pull-up/pull-down resistors, and input mode. 

\Paragraph{Transmit Signal}
Given that an \ac{RF} sensitivity on a particular combination of a reception path and configuration exists, the incident \ac{RF} waveform must match in terms of carrier frequency and signal power, where we assume $x(t) = \sqrt{P_{Tx}}\ m(t)\ \cos(2\pi f_{Tx} t)$ as the stimulus signal, where $P_{Tx}$ is the transmit power, $m(t)$ is an amplitude-modulation, and $f_{Tx}$ denotes the carrier frequency. %

\subsection{Experimental Setup}

\begin{figure}
    \centering
    \includegraphics[width=0.9\linewidth]{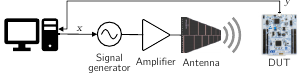}
    \caption{Illustration of our experimental setup.}
    \label{fig:exp_setup}
\end{figure}

To systematically search for \ac{RF} sensitivities, we built a controlled measurement setup as illustrated in~\autoref{fig:exp_setup}. A control computer orchestrates the experiment, configuring both the transmitter and the \ac{DUT} receiver, triggering measurements, and collecting \ac{ADC} samples for analysis. We conduct the experiments in an office and in the basement of a typical office building with ambient radio activity by nearby wireless networks such as \mbox{Wi-Fi}. In both settings, concrete walls and silver-coated windows provide sufficient shielding to avoid interference with radios outside the building.

As a programmable \ac{RF} signal source, we employ a SignalHound VSG60 vector signal generator to precisely control the signal power $P_{Tx}$ from \SI{-40}{dBm} to \SI{-7}{dBm} and vary the signal frequency $f_{Tx}$ between \SI{200}{MHz} and \SI{1000}{MHz}. The generator output is amplified by a Mini-Circuits ZHL-20W-13SW+ power amplifier providing approximately \SI{50}{dB} of gain and enabling transmit powers up to \SI{43}{dBm} ($\approx$~\SI{20}{W}). The signal is emitted by an RFspace LPDA-max antenna that has a gain of approximately \SI{6.5}{dBi}. Unless stated otherwise, the antenna is placed at a distance of \SI{1}{\m} from the \ac{DUT}. 

The \ac{DUT} can be any embedded device with an \ac{ADC}, a serial communication interface, and that we can program the firmware on. Unless otherwise noted, the \ac{DUT} is an off-the-shelf Nucleo~G474RE development board featuring an STM32~\ac{MCU}. Beyond this reference platform, we further evaluate 13~other devices such as commercially available cryptocurrency hardware wallets based on seven other \acp{MCU}~(see~\autoref{sec:casestudies}). The available reception paths vary by device, \eg, depending on the number of \ac{GPIO} pins, internal \ac{ADC} channels, and select on-chip signals. For the path configurations, we consider $64$ combinations of the \ac{GPIO} parameters mode (input, output, analog, alternate function), pupd (pull-up / pull-down / none / reserved), value (high, low), output type (open-drain / push-pull). We also configure the \ac{ADC} parameters, including the total number of samples per reading, the sample rate, and oversampling.

The host computer communicates with the \ac{DUT} through a serial interface to configure the \ac{MCU} to a specific combination of reception path and path configuration, trigger \ac{ADC} sampling, and retrieve \ac{ADC} samples for offline processing. This automated workflow enables exhaustive exploration across reception paths, pin configurations, and transmit signal parameters, forming the foundation for our systematic identification of \ac{RF} sensitivities and their subsequent exploitation.

\subsection{Finding RF Sensitivities}
\label{sec:finding_rf}
\begin{figure}[htb]
    \centering
    \includegraphics[width=0.9\linewidth]{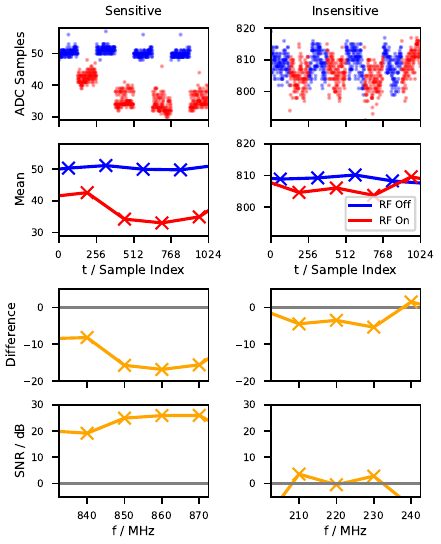}
    \caption{Processing for \ac{RF} sensitivity testing. First row: raw~\ac{ADC} samples from the device over time while the transmitter is switched on and off and changing frequency. Second row: average over sample blocks. Third row: Difference of block averages between on and off samples over frequency. Fourth row: \ac{SNR} estimation over frequency.}
    \label{fig:adc_data_processing}
\end{figure}

We take a phenomenological, measurement-driven approach, empirically characterizing how embedded devices respond to controlled \ac{RF} stimuli (unmodulated carrier signals at full transmit power). Our \ac{RF} sensitivity search proceeds exhaustively: We first select a specific reception path–configuration pair on the \ac{DUT} and then collect a block of $N$~\ac{ADC} samples with the \ac{RF} generator off, followed by another block with the generator on. This process is repeated while stepping the carrier frequency, producing a frequency sweep for each possible combination of reception path and configuration.

\begin{figure}[htb]
    \centering
    \includegraphics[width=0.8\linewidth]{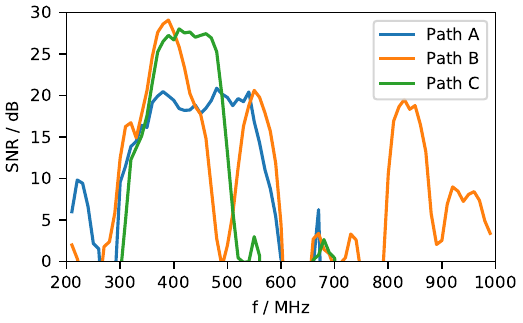}
    \caption{\ac{SNR} over frequency for three different reception paths on the Nucleo-G474RE board.}
    \label{fig:per_path_sensitivity}
    \Description{}
\end{figure}

Using this method, we characterize all reception paths of the Nucleo-G474RE development board and the Passport hardware wallet, each tested under $64$ distinct path configurations and $81$ evenly spaced carrier frequencies between \SI{200}{\MHz} and \SI{1000}{\MHz}. For each combination, we collect $32$~ADC samples for both the RF-on and RF-off states. To determine whether a given configuration exhibits sensitivity, we apply the analysis pipeline illustrated in~\autoref{fig:adc_data_processing}. Specifically, we compute the mean of each ADC sample block and evaluate the mean difference between the RF-on and RF-off states across frequency steps. This differential trace captures the device’s response to incident \ac{RF} energy. Finally, we estimate the corresponding \ac{SNR} using the noise variance observed in the RF-off measurements. The frequency-dependent \ac{SNR} estimates of three representative and particularly sensitive path-configuration combinations are shown in~\autoref{fig:per_path_sensitivity}.

\begin{figure}[htb]
    \centering
    \includegraphics[width=1\linewidth]{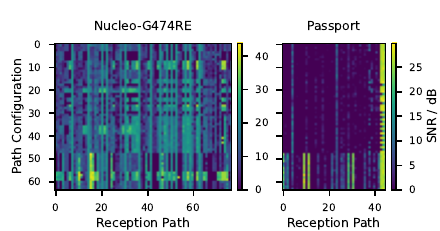}
    \caption{Full sensitivity testing experiment results for the Nucleo-G474RE \tablecirclesmall{10} and the Passport \tablecirclesmall{5} boards, showing the peak \ac{SNR} over frequency for combinations of reception path and \ac{GPIO} configuration.}
    \label{fig:gpio_test}
    \Description{}
\end{figure}

\autoref{fig:gpio_test} shows the peak~\ac{SNR} across frequency for each path-configuration combination on the Nucleo-G474RE board and the Passport hardware wallet. Both devices exhibit pronounced \ac{RF} sensitivities, with several configurations exceeding \SI{20}{\dB}. The Nucleo-G474RE generally shows stronger and more frequent responses, but in both devices, the reception path and configuration strongly influence the observed \ac{SNR}. Notably, in the lower sections of both heatmaps, elevated sensitivity occurs where the \ac{GPIO} is set to analog mode. On the Nucleo-G474RE, additional peaks appear around configuration indices 10, 20, and 35, where pull-down or pull-up resistors are activated, also yielding higher \ac{SNR}.

\begin{figure}[htb]
    \centering
    \includegraphics[width=1\linewidth]{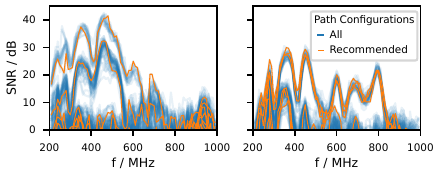}
    \caption{\ac{SNR} over frequency on two reception paths for all $64$~possible path configurations and the recommended path configurations.}
    \label{fig:all_vs_recommended_path_configurations}
    \Description{}
\end{figure}

Manual inspection of all $64$ configurations reveals significant redundancy: many configurations produce largely similar responses. Our analysis indicates that the primary factors governing \ac{RF} sensitivity are the \ac{GPIO} input mode and the pull-up/pull-down resistor settings. Based on these insights, we distilled a subset of 8 recommended path configurations (see~\autoref{sec:recommended_path_configurations}), sufficient to characterize most sensitivity patterns. \autoref{fig:all_vs_recommended_path_configurations} illustrates this for two reception paths on the Nucleo-G474RE: the plot compares sensitivity spectra across all configurations and highlights the recommended subset, demonstrating that these selected configurations effectively capture all distinct classes of \ac{RF} sensitivity. Apart from the path-configuration combintion, we found that the \ac{ADC}-based sampling configuration affects the \ac{SNR}. As expected, oversampling can be used to enhance \ac{ADC} resolution and mitigate noise, thereby improving \ac{SNR} (see Figure~23, extended version~\cite{staatAirgapExtended} Appendix~G).

\section{How Common are RF Sensitivities in the Wild?}
\label{sec:casestudies}
So far, we have outlined and demonstrated our methodology for identifying \ac{RF} sensitivities on two representative devices. On both the Nucleo-G474RE board and the Passport hardware wallet, we found clear sensitivities with \acp{SNR} sufficiently high to enable covert wireless communication. A natural next question is whether these are device-specific isolated observations or whether such \ac{RF} sensitivities are a more widespread phenomenon among embedded devices.

\begin{figure}
    \centering
    \includegraphics[width=0.9\linewidth]{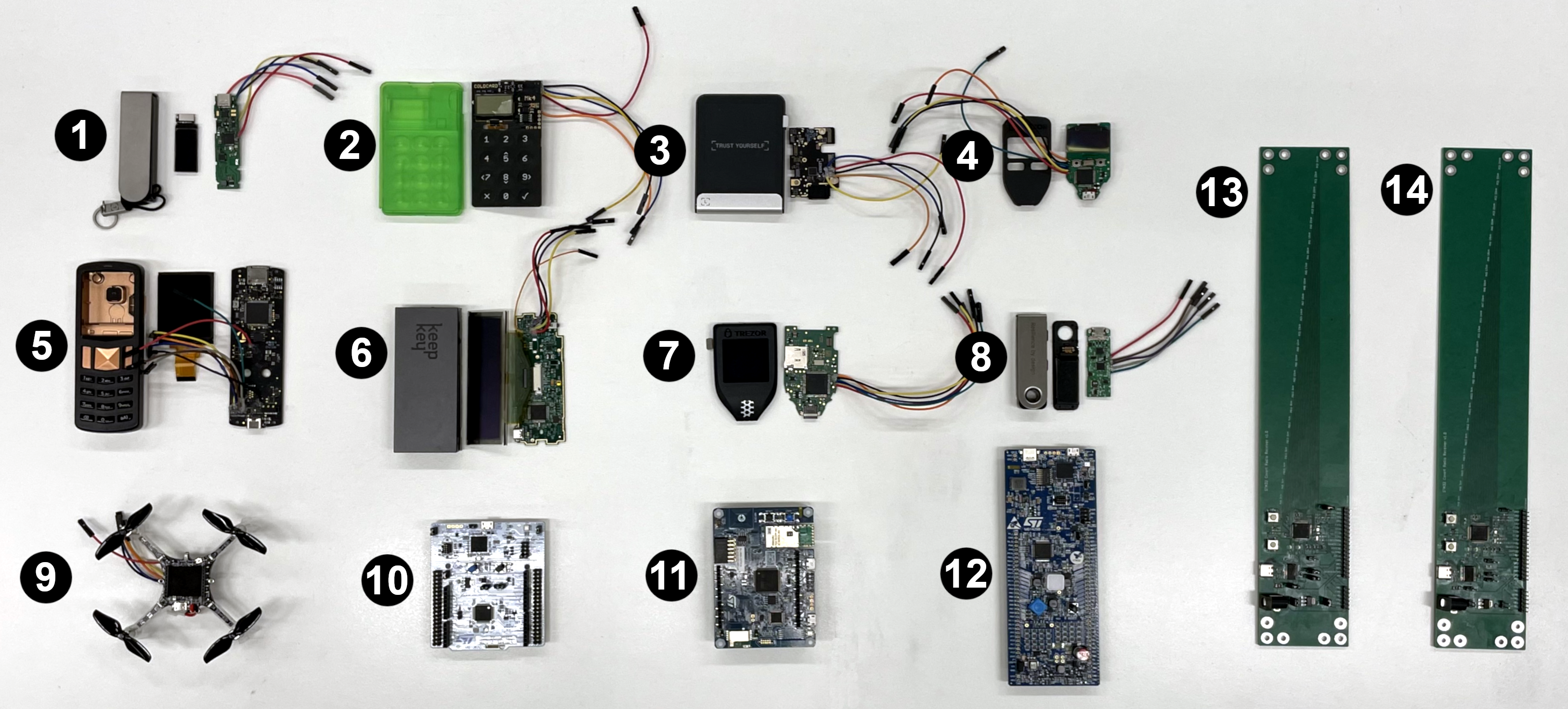}
    \caption{All tested devices.}
    \label{fig:all_devices}
\end{figure}

To investigate this, we ran our sensitivity discovery method on a total of 14~devices, comprising 12~commercial products and 2~custom devices. Our selection includes eight cryptocurrency hardware wallets (some of which are explicitly advertised as air-gapped), an electronics hobbyist drone, multiple \ac{MCU} development boards, and two custom \acp{PCB} (see~\autoref{sec:countermeasures} for details on the custom designs). The tested devices are depicted in~\autoref{fig:all_devices}. Based on the insights gained from testing the Nucleo-G474RE and the Passport, we restricted our search across path configurations to the eight~recommended choices, varying the \ac{GPIO} input mode and pull-up/down configurations.

For the cryptocurrency wallets, we de-soldered the original \acp{MCU} and replaced them with identical, blank parts to obtain full firmware control over the hardware platform. Our goal in doing so was strictly to evaluate \ac{RF} reception characteristics -- not to target secure boot or similar protection mechanisms. Attacks to break such mechanisms have been studied extensively in prior work~\cite{roth2018ccc, cuiBADFETDefeatingModern2017, wouters2022blackhat} and are orthogonal to our contribution. 

We conducted all experiments on the bare \acp{PCB} of the devices, isolating the fundamental receive mechanism from potential effects of device enclosures. To realize automated device control and data streaming at scale, we relied on wired USB connections, so our results reflect the sensitivity of the device in conjunction with the cable. We show in \autoref{sec:wiring_impact} that sensitivities persist without the cable and that cabled measurements are representative of air-gapped ones. Regarding enclosures, we show in \autoref{sec:countermeasures} that a plastic enclosure does not prevent reception, while a metallic enclosure can mitigate it. In future work, we will further investigate the influence of enclosures and electronic peripherals such as displays and cables.

\autoref{tab:case_study_devices} summarizes the results for all tested devices. For each device (\cf the numbering in~\autoref{fig:all_devices}), we list the \ac{MCU} model, the number of sensitive versus total reception paths, the most sensitive path and its configuration (if applicable), the frequency range of observed sensitivity, and the peak measured \ac{SNR}. The central finding is clear: \emph{every} tested device exhibited measurable \ac{RF} sensitivity. This confirms that the phenomenon is not tied to a particular hardware design, but rather a recurring effect that manifests broadly across embedded platforms. In several cases, the incident \ac{RF} signal induced large shifts in \ac{ADC} traces that otherwise show no variance. For such cases, the \acp{SNR} could not be reliably estimated and are therefore reported as \enquote{high}. At the same time, the results reveal pronounced variation across device \acp{PCB}: while some, such as the Ledger Nano S Plus and Ledger Flex, exhibit only weak sensitivity, others, such as the COLDCARD™ Mk4, the Passport, and our custom board, show stronger and more sensitivities.

\begin{table*}
\caption{Sensitivity results for all tested devices.}
\label{tab:case_study_devices}
\centering
{\fontsize{6}{7}\selectfont\def\arraystretch{2}\tabcolsep=3pt

\begin{tabular}{llrlllrr}
	\toprule
	\textbf{Device}    & \textbf{Microcontroller} & \makecell[l]{\textbf{Sensitive}\\\textbf{Paths}} & \textbf{Best Reception Path} & \makecell[l]{\textbf{Path}\\\textbf{Configuration}} & \textbf{Frequency [MHz]} & \textbf{Peak SNR [dB]} \\ \toprule
	
	\tablecircle{1} Ledger Nano X      & STM32WB35CC              & 18 / 44 & ADC Channel VREFINT (single-ended)                     & -                     &        300 - 400, 450 - 650, 735                  &    24               \\ %

    \tablecircle{2} COLDCARD™ Mk4       & STM32L4S5VI              & 5 / 34 & ADC Channel 11, differential-ended with GPIOs PA6 and PA7                    & \texttt{analog, pull-down, set}            &   320 - 400, 420 - 530, 660 - 680                       &      high               \\ %

	\tablecircle{3} Ledger Flex      & STM32WB35CC              & 2 / 44 & ADC Channel VREFINT (single-ended)                     & -                     & 500 - 600                         &    12               \\ %

	\tablecircle{4} Trezor Model One   & STM32F205RE              & 5 / 43 & ADC2 Channel 7 with GPIO PA7 & \texttt{analog, pull-up, reset}                    &  250 - 525 &                15 \\ %

    \tablecircle{5} Passport           & STM32H753VI              & 24 / 45 & ADC3 Channel VBAT (single-ended)                     &    -                  &                   200 - 250, 280, 320 - 1000       &       34            \\

	\tablecircle{6} KeepKey            & STM32F205RG              & 5 / 43 & ADC1 Channel VREFINT   &    -                  &   350 - 400, 600 - 1000        &                  16   \\

	\tablecircle{7} Trezor Model T     & STM32F427VI              & 13 / 57 & ADC3 Channel 17 (reserved) &     -                 &            300 - 600              &      14             \\ %

	\tablecircle{8} Ledger Nano S Plus & STM32F042K6              & 5 / 13 &           ADC Channel 4 with GPIO PA4 & \texttt{input, pull-down, set}  &       325 - 375, 600                &                 11 \\ %

    \midrule

	\tablecircle{9} CrazyFlie 2.1+     & STM32F405RG              & 19 / 57 &    ADC3 Channel 11 with GPIO PC1                  &       \texttt{analog, pull-up, reset}               &         200 - 1000                 &            62       \\

    \tablecircle{10} Nucleo-G474RE & STM32G474RE & 53 / 87 & ADC2 connected to OPAMP5 with GPIO PC3 & \texttt{AF, pull-down, set} & 200 - 1000 & 33  \\ %

	\tablecircle{11} \makecell[l]{STM32L4+\\Discovery IoT Kit} & STM32L4S5VI & 9 / 44 & ADC Channel 12 with GPIO PA7 & \texttt{input, pull-up, reset} & 400 - 725, 820 - 850 & 19  \\

	\tablecircle{12} \makecell[l]{STM32G474RE\\Discovery Kit} & STM32G474RE & 45 / 87 & ADC1 Channel 8 with GPIO PC2 & \texttt{analog, pull-down, set} & 250 - 1000 & 37  \\ %

	\tablecircle{13} Custom PCB & STM32G474RE & 87 / 87 & ADC2 Channel 6 with GPIO PC0 & \texttt{analog, pull-up, reset} & 200 - 1000 & high  \\ %

   	\tablecircle{14} \makecell[l]{Custom PCB\\/w ground plane} & STM32G474RE & 53 / 87 & ADC2 connected to OPAMP2 with GPIO PB0 & \texttt{input, pull-up, reset} & 200 - 1000 & high  \\ %

 \bottomrule
	\end{tabular}
}
\end{table*}

\section{Sensitivity Characterization}
\label{sec:sensitivity_experiments}

In this section, we systematically characterize the \ac{RF} sensitivities of the Nucleo-G474RE board, analyzing their stability over time, device dependence, and sensitivity to orientation.

\subsection{Dependence on Signal Power}

A critical question is whether the observed \ac{RF} sensitivities could enable wireless communication over larger distances. To examine this, we repeated the \ac{SNR} estimation for the sensitivities shown in~\autoref{fig:per_path_sensitivity} at fixed frequencies while progressively reducing the transmit power. Using the known antenna-device distance, antenna gain, frequency, and transmission power, we estimate the signal arrival power at the \ac{DUT} as the power that would be delivered into a matched \SI{50}{\ohm} load at the receiver location. As our \acp{DUT} present no defined input impedance, this estimate serves as a reference value rather than the power actually coupled into the device. Absolute coupling efficiency is therefore not captured, and we use the metric only for relative comparisons, \eg, across frequency, distance, and configuration.

\begin{figure}
    \centering
    \includegraphics[width=0.82\linewidth]{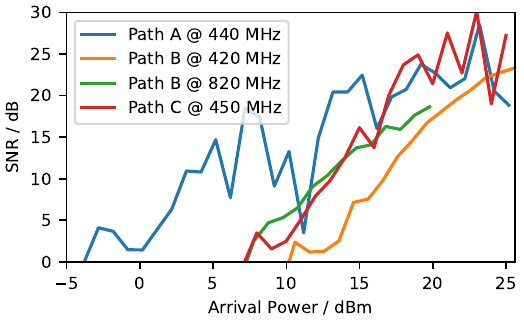}
    \caption{\ac{SNR} over estimated arrival signal power (50 \si{\ohm}-equivalent) for three reception paths.}
    \label{fig:snr_against_power}
    \Description{}
\end{figure}

\autoref{fig:snr_against_power} plots the resulting \ac{SNR} as a function of the estimated arrival power, showing that signal powers as low as \SI{0}{dBm} remain detectable. Interestingly, the \ac{SNR} increases superlinearly with signal power, indicating the presence of nonlinear effects which we attribute to the unconventional reception mechanism. These results demonstrate that the observed sensitivities persist at lower signal powers rather than disappearing abruptly. In fact, as we will see in data transmission experiments in~\autoref{sec:communication}, reception on path~B extends to even weaker signal levels.

\subsection{Stability Over Time}

\begin{figure}[htb]
    \centering
    \includegraphics[width=0.9\linewidth]{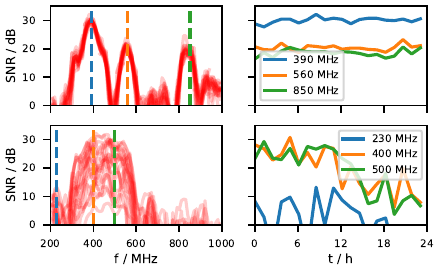}
    \caption{Sensitivity assessment over time for two separate reception paths~(top: path~B, bottom: path~A) over frequency~(left) and time~(right).}
    \label{fig:stability_over_time}
    \Description{}
\end{figure}

To evaluate the temporal stability of the identified \ac{RF} sensitivities, we repeat the sensitivity characterization periodically over a \SI{24}{h} period. In each iteration, we measure the device’s response while alternately enabling and disabling the \ac{RF} transmit signal. As shown in~\autoref{fig:stability_over_time} for two representative reception paths, the frequency-dependent sensitivity patterns remain largely consistent over time. The first path (top) exhibits high stability with nearly unchanged sensitive frequency regions, whereas the second path (bottom) shows moderate drift and both short- and long-term fluctuations, evident from the \ac{SNR} degradation after approximately \SI{15}{h}. Overall, these results indicate that the device’s unintended \ac{RF} reception behavior is persistent and only mildly affected by temporal variations.

\subsection{Inter-Device Variation}
\label{sec:inter_device}

So far, we have established that \ac{RF} sensitivities on a given device can exist across several peripheral configurations and remain stable over time. In a realistic airgap infiltration scenario, however, the attacker may not have direct access to the target hardware for characterization. While such access might still be possible—through brief onsite presence or a supply-chain compromise—it is considerably more practical for the attacker to obtain an identical device model and characterize its reception sensitivities in advance~\cite{yangReThinkRevealThreat2025}. The key question, therefore, is whether \ac{RF} sensitivities discovered on one sample of a device type are transferable to other samples of the same type.

\begin{figure}
    \centering
    \includegraphics[width=1\linewidth]{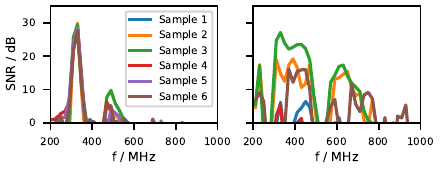}
    \caption{\ac{SNR} over frequency for two distinct reception paths on six different Nucleo-G474RE boards, showing that sensitives are repeatable.}
    \label{fig:inter_device_reproducibility}
    \Description{}
\end{figure}

To answer this, we examine six Nucleo-G474RE boards under identical test conditions, ensuring consistent placement, orientation, and environmental conditions. \autoref{fig:inter_device_reproducibility} shows the measured \ac{SNR} over frequency for two representative reception paths across all six samples. On the left, we observe near-perfect agreement between devices: the sensitive frequency regions and \ac{SNR} results align closely across all samples. On the right, however, only a subset of devices exhibit notable sensitivity, while others (\eg, sample~5) show none. Among the sensitive samples, the spectral shape and frequency range of the sensitivities remain largely consistent. In our sensitivity assessments, testing all reception paths with the eight recommended GPIO~configurations, we found $55$~combinations with \ac{RF} sensitivity on all six samples, see \autoref{sec:inter_device_qty} for a more detailed overview. In short, sensitivities are repeatable across device instances: an attacker can reliably identify device-repeatable sensitivities on surrogate units and then transfer those findings to the target device. %

\subsection{Device Orientation}

\begin{figure}
    \centering
    \includegraphics[width=0.9\linewidth, trim={0 6mm 0 0},clip]{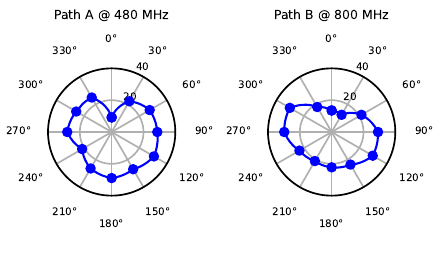}
    \caption{\acp{SNR} at a single frequency for two reception paths over device rotation (in the $xz$ plane).}
    \label{fig:snr_vs_rotation}
    \Description{}
\end{figure}

To assess whether device orientation influences \ac{RF} sensitivity, we measured the \ac{SNR} over frequency for two reception paths while systematically changing the device orientation. Specifically, we measured the \ac{SNR} at a fixed frequency while rotating the device in the $xz$ plane. \autoref{fig:snr_vs_rotation} shows the results over the \ac{DUT} angle with respect to the transmitter. Here, we observe slight directionality: Path~A shows reduced sensitivity near \SI{0}{\degree} but remains otherwise stable, whereas path~B exhibits stronger responses around \SI{90}{\degree} and \SI{270}{\degree}. Further, we found that placing the device in other planes affects the reception sensitivity but does not eliminate the sensitivity phenomenon (see Figure~24, extended version~\cite{staatAirgapExtended} Appendix~H).

\subsection{Impact of Wired Connections}
\label{sec:wiring_impact}

We verify that the observed \ac{RF} sensitivities originate from the device itself rather than from external coupling paths. Having ruled out a common-ground effect between \ac{DUT} and transmitter (see Figure~22, extended version~\cite{staatAirgapExtended} Appendix~F), we turn to the USB cable used for power, control, and data transfer. Such cables are part of the physical system and may legitimately act as antennas, but isolating their contribution is necessary to understand the reception mechanism and assess generality. 

\begin{figure}
    \centering
    \includegraphics[width=.9\linewidth]{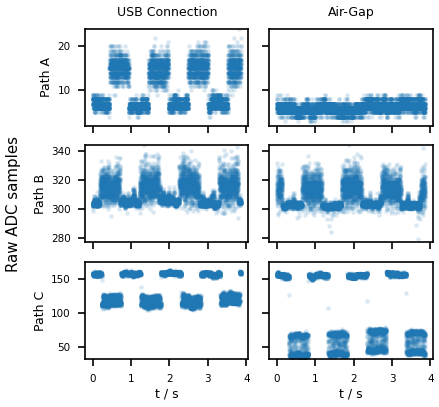}
    \caption{Raw \ac{ADC} recordings with USB connection (left) and fully air-gapped operation (right) while turning the \ac{RF} transmission on and off every \SI{0.5}{\m}.}
    \label{fig:airgap_test}
    \Description{}
\end{figure}

We toggle the transmitter every \SI{0.5}{\second} under two conditions: with the USB cable attached, and fully air-gapped, with the \ac{DUT} battery-powered and \ac{ADC} recording triggered by a physical button press, samples being retrieved after reconnecting the serial interface. \autoref{fig:airgap_test} compares raw \ac{ADC} traces. Path~A loses all sensitivity without the cable, indicating the cable acted as antenna. Path~B is unchanged and Path~C is more sensitive when air-gapped. \ac{RF} sensitivity is thus attributable to intrinsic board-level behavior, though external connections can contribute.

To isolate the cable's contribution more broadly, we applied the sensitivity testing method of \autoref{sec:finding_rf} to a reduced sweep: one frequency, 11 reception paths, five GPIO configurations, measured cabled and then battery-powered. \autoref{fig:usb_vs_battery} in the Appendix plots the mean differences against each other. The two conditions track closely, indicating that USB-cabled sensitivity testing is representative of truly air-gapped conditions.

\section{Reception of Communication Signals}
\label{sec:communication}

We now explore how the identified \ac{RF} sensitivities can be exploited for wireless data reception. We first characterize \ac{BER} and data rate performance, and then demonstrate a full-blown data reception under real-world conditions over a distance of \SI{20}{m}.

The \ac{RF} sensitivities characterized in the previous sections and wireless data reception rely on the exact same physical mechanism. \autoref{tab:case_study_devices} indicates sufficiently high \acp{SNR} for all tested devices. We therefore expect generality across devices and implement full data reception on only one exemplary device accordingly.

\subsection{Modulation Scheme}
Given that the identified \ac{RF} sensitivities allow reliable distinction between the presence and absence of a carrier, the most natural choice for digital modulation is \ac{OOK}, where the transmitter is enabled for a `\texttt{1}' and disabled for a `\texttt{0}'. In all subsequent experiments, we employ \ac{OOK} with rectangular pulse shaping. However, since the received signal strength scales with transmit power (see~\autoref{fig:snr_against_power}), higher-order \ac{ASK}, \eg, with four symbols, is in principle feasible.

\subsection{Ideal Receiver Bit Error Rates}

To assess whether the observed \ac{RF} sensitivities can be leveraged for wireless data reception, we conducted the following experiment. For a single reception path, we recorded $10{,}000$~\ac{ADC} sample blocks while the \ac{RF} signal generator was randomly either switched on or off, effectively emulating an \ac{OOK} transmission. We assume perfect transmitter–receiver synchronization (\ie, we collect $127$ samples per bit) and neglect transient effects between the on and off states. As a simple demodulator, we apply a moving average filter to track slow variations in the \ac{DC} offset, using its output as a decision threshold. Blocks with an average above this threshold are decoded as a `\texttt{1}', and those below as a `\texttt{0}'. To prevent biasing of the decision threshold, long identical bit runs can be prevented through standard line coding and interleaving techniques. Comparing the resulting bitstream with the known transmission sequence allows us to compute the corresponding \ac{BER}.

\begin{figure}
    \centering
    \includegraphics[width=1.0\linewidth]{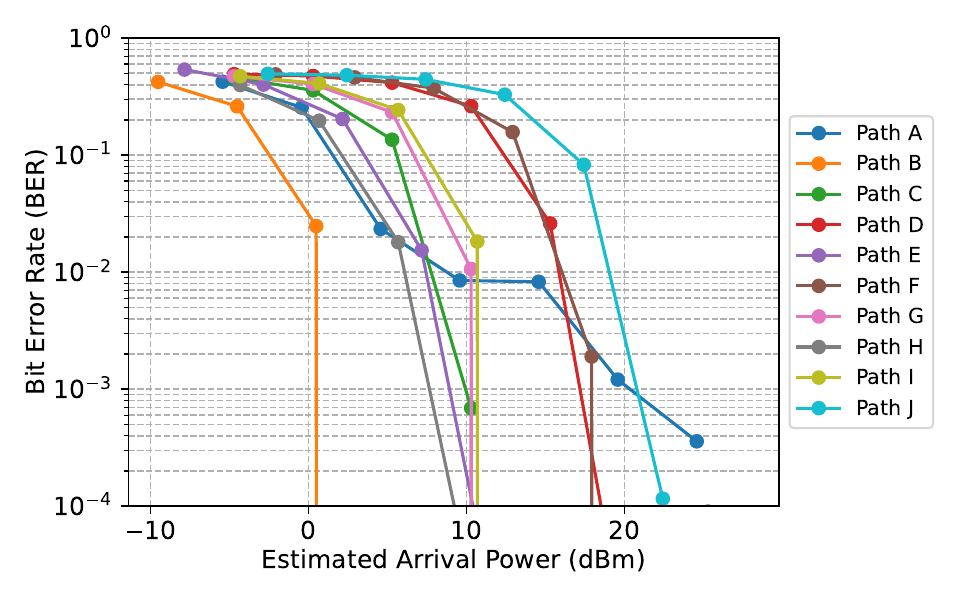}
    \caption{\ac{BER} for several reception paths on the Nucleo-G474RE over estimated arrival power (\SI{50}{\ohm}-equivalent) under ideal synchronization.}
    \label{fig:ber_vs_tx_power}
    \Description{}
\end{figure}

We repeated this experiment for ten different reception paths and multiple transmit powers. Using the known antenna distance and carrier frequency, we estimate the power arriving at the device (Nucleo~G474RE) and plot the resulting \ac{BER} curves in~\autoref{fig:ber_vs_tx_power}. The results reveal two key findings: ($i$)~reception path~B achieves the best performance, with a \ac{BER} below \SI{2}{\percent} at an arrival power of \SI{0}{dBm} ($\approx$~\SI{1}{\mW}), and ($ii$)~the achievable sensitivity strongly depends on the chosen reception path. Still, 7 of 10 paths reach a \ac{BER} below \SI{1}{\percent} when the arrival signal power is \SI{10}{dBm} ($\approx$~\SI{10}{\mW}). The non-monotonic \ac{BER} curve of path~A is due to strong noise and drift on the path. Overall, this experiment demonstrates that the identified \ac{RF} sensitivities can indeed support reliable digital communication.

\Paragraph{Bandwidth Considerations}
In the previous experiment, we ignored transient effects introduced by switching the \ac{RF} carrier on and off. For reliable \ac{OOK} reception, however, the underlying \ac{RF} sensitivity must provide sufficient bandwidth so that bit transitions remain distinguishable. To evaluate such effects qualitatively, we configure the Nucleo-G474RE board to reception path~B and transmit \ac{OOK} signals with alternating `\texttt{0}'/`\texttt{1}' patterns at data rates between \SI{500}{bps} and \SI{100000}{bps}. The \ac{ADC} sampling rate is adjusted to ensure adequate oversampling for all rates.

\begin{figure}
     \centering
     \includegraphics[width=0.9\linewidth]{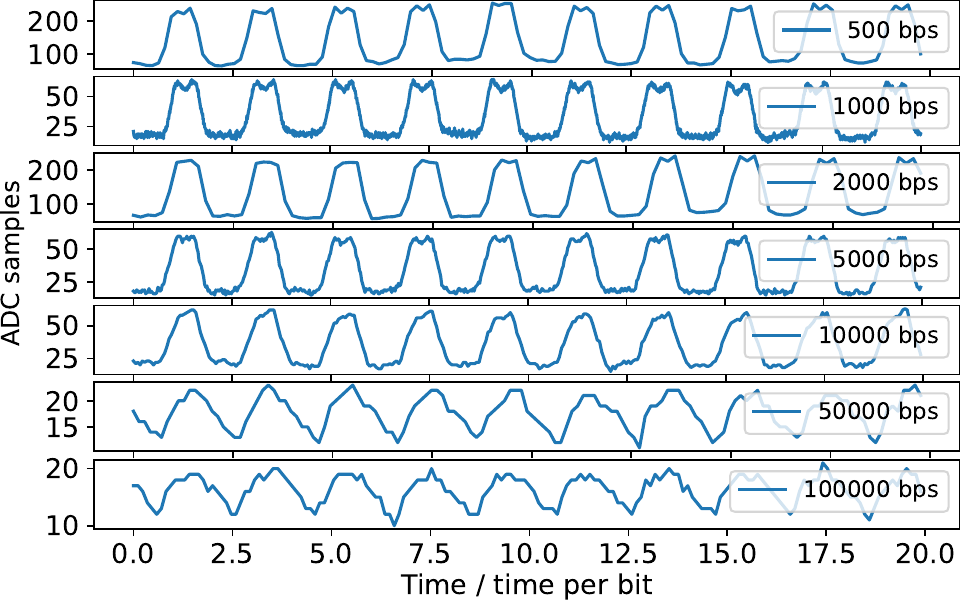}
     \caption{Illustration of receive signal waveform over increasing symbol rates, showing a decreasing slew rate with increasing symbol rate.}
     \label{fig:transients}
     \Description{}
\end{figure}

\autoref{fig:transients} shows $20$~received bits per rate and we observe that edge steepness reduces with higher data rates, causing increasing smearing between adjacent bits. This result suggests a limited bandwidth of the reception path. Nonetheless, even at \SI{100}{\kilo bps}, the bit pattern remains discernible. 

\subsection{Real-World Data Reception}

\begin{figure*}%
    \begin{minipage}[c]{0.38\columnwidth}
    \begin{subfigure}{1\columnwidth}
        \centering
        \includegraphics[width=\columnwidth]{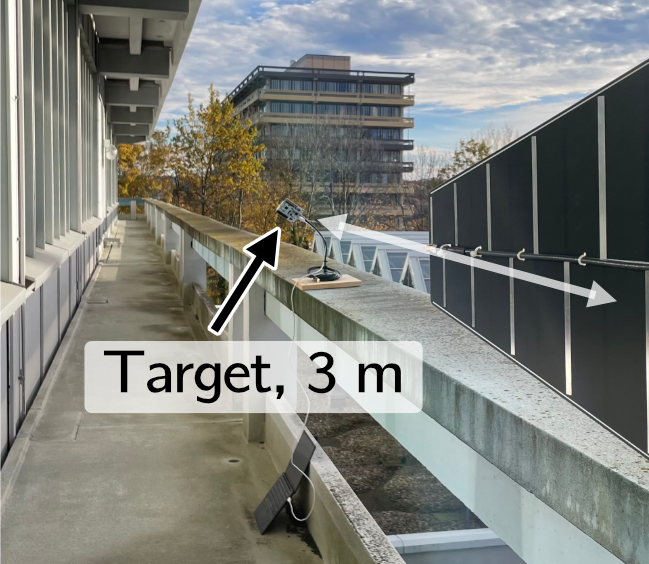}
        \caption{}
        \label{fig:close_transmission_img}
    \end{subfigure}
    \end{minipage}
    \begin{minipage}[c]{0.55\columnwidth}
    \begin{subfigure}{1\columnwidth}
        \centering
        \includegraphics[width=\columnwidth]{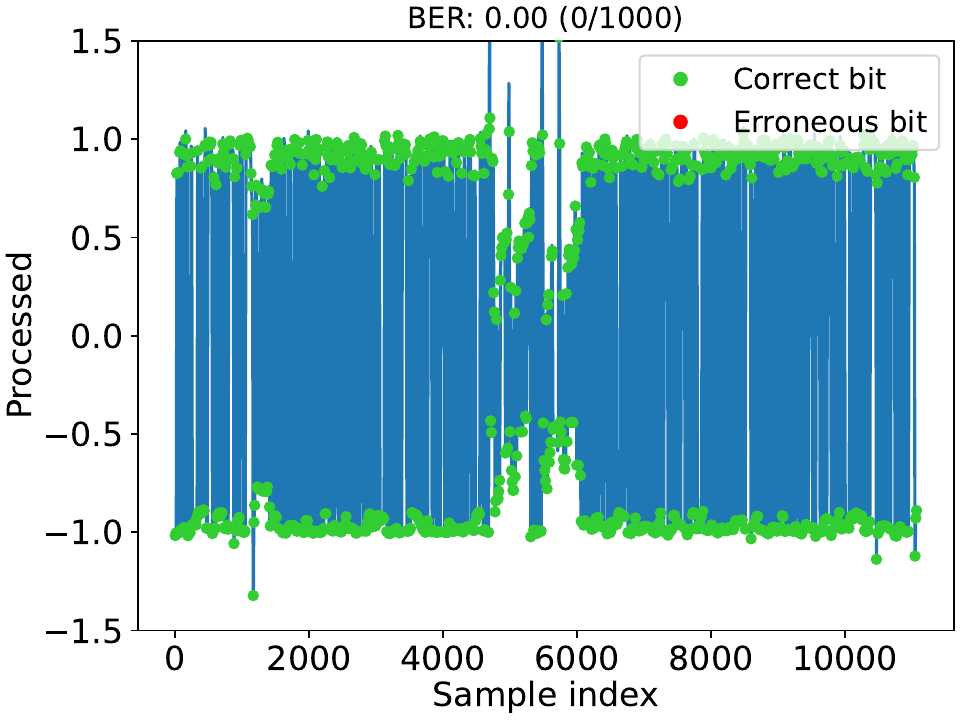}
        \caption{}
        \label{fig:close_transmission_signals}
    \end{subfigure}
    \end{minipage}
    \begin{minipage}[c]{0.55\columnwidth}
    \begin{subfigure}{1\columnwidth}
        \centering
        \includegraphics[width=\columnwidth]{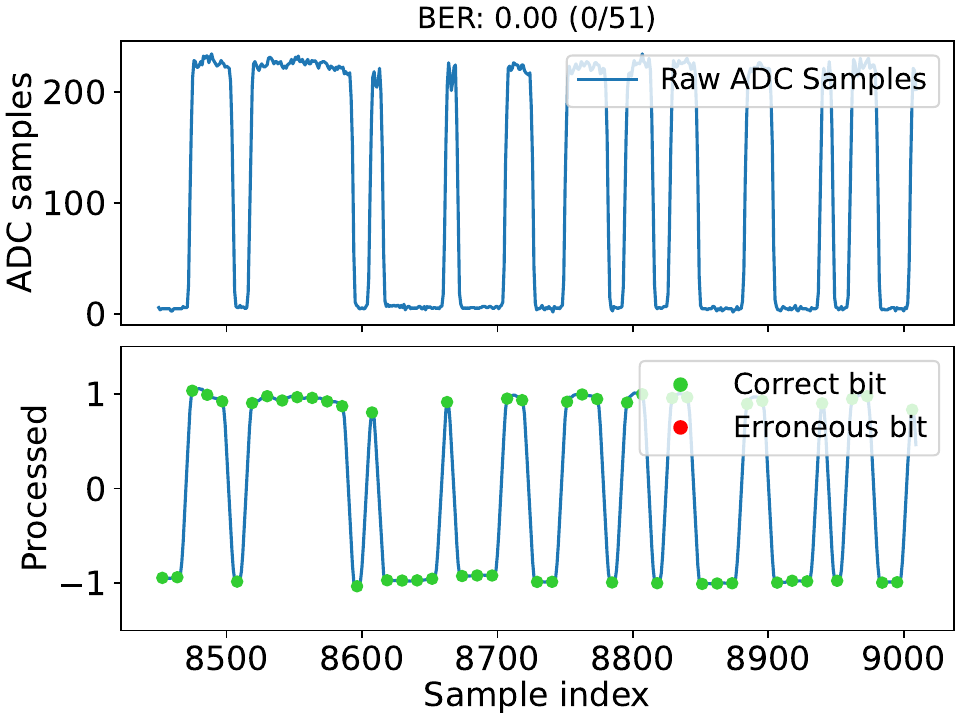}
        \caption{}
        \label{fig:close_transmission_signals_macro_perfect}
    \end{subfigure}
    \end{minipage}
    \begin{minipage}[c]{0.55\columnwidth}
    \begin{subfigure}{1\columnwidth}
        \centering
        \includegraphics[width=\columnwidth]{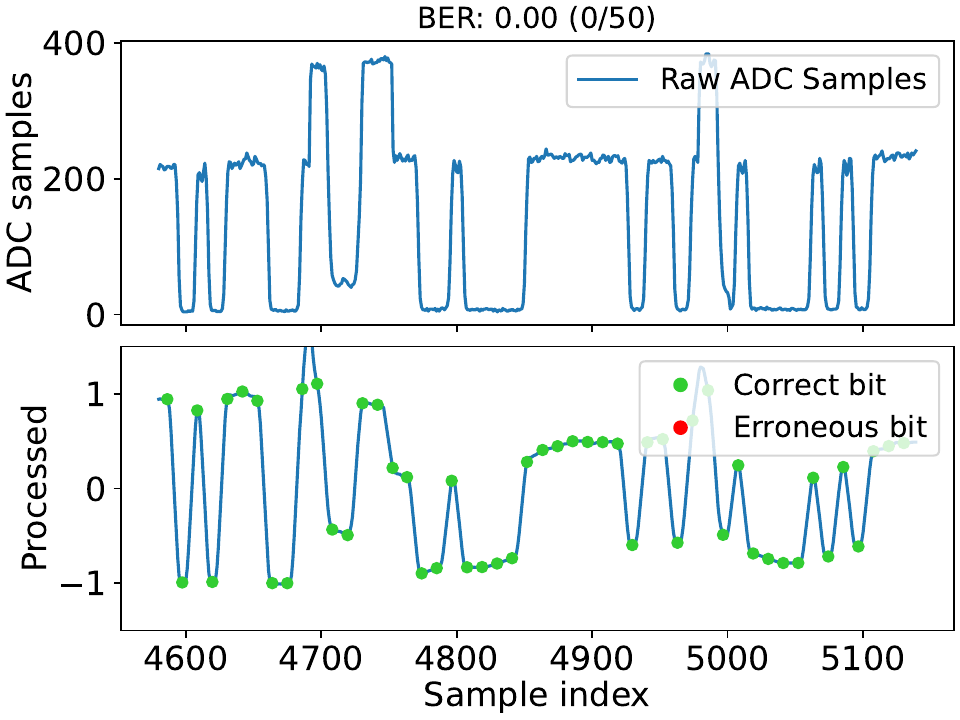}
        \caption{}
        \label{fig:close_transmission_signals_macro_imperfect}
    \end{subfigure}
    \end{minipage}\\
    
    \begin{minipage}[c]{0.38\columnwidth}
    \begin{subfigure}{1\columnwidth}
        \centering
        \includegraphics[width=\columnwidth]{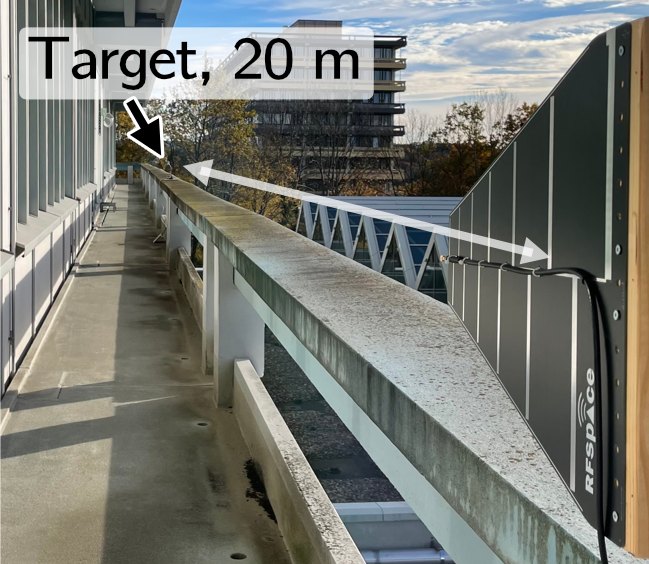}
        \caption{}
        \label{fig:far_transmission_img}
    \end{subfigure}
    \end{minipage}
    \begin{minipage}[c]{0.55\columnwidth}
    \begin{subfigure}{1\columnwidth}
        \centering
        \includegraphics[width=\columnwidth]{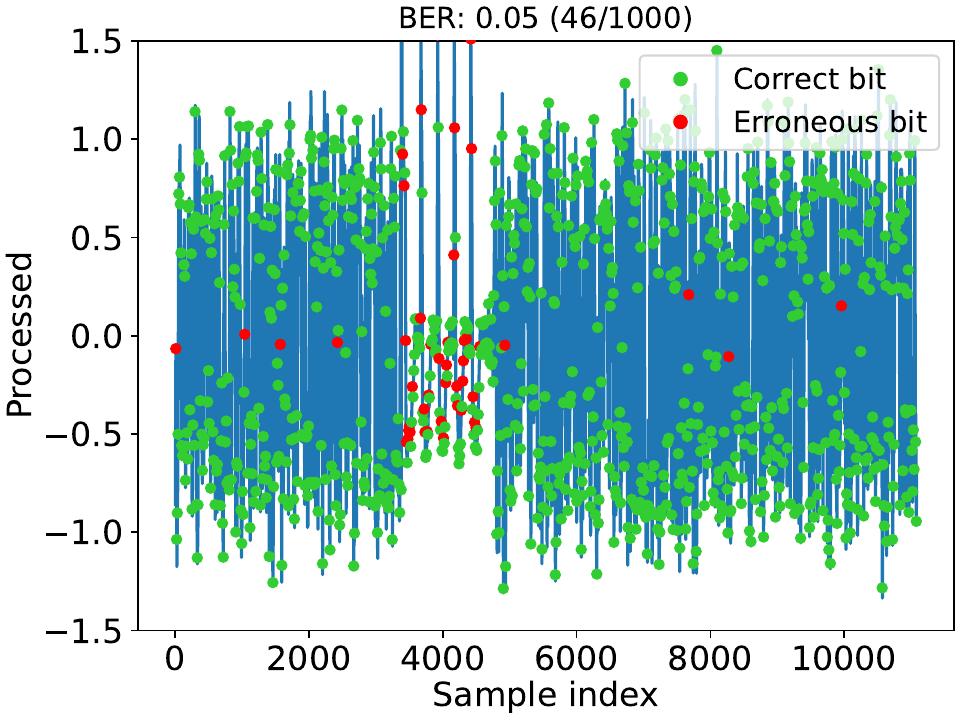}
        \caption{}
        \label{fig:far_transmission_signals}
    \end{subfigure}
    \end{minipage}
    \begin{minipage}[c]{0.55\columnwidth}
    \begin{subfigure}{1\columnwidth}
        \centering
        \includegraphics[width=\columnwidth]{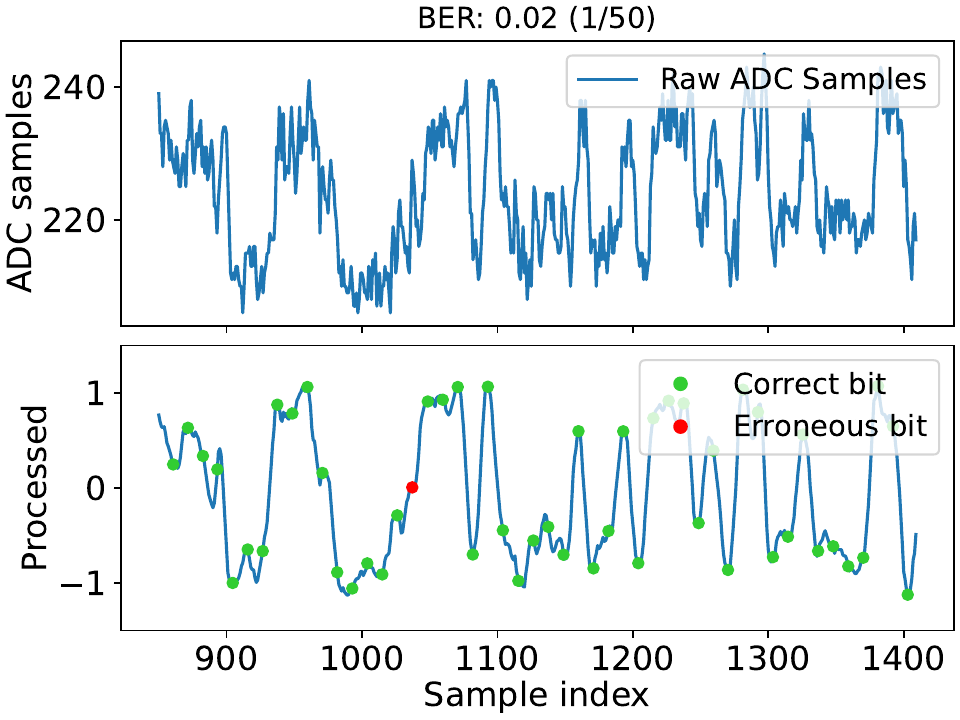}
        \caption{}
        \label{fig:far_transmission_signals_macro_perfect}
    \end{subfigure}
    \end{minipage}
    \begin{minipage}[c]{0.55\columnwidth}
    \begin{subfigure}{1\columnwidth}
        \centering
        \includegraphics[width=\columnwidth]{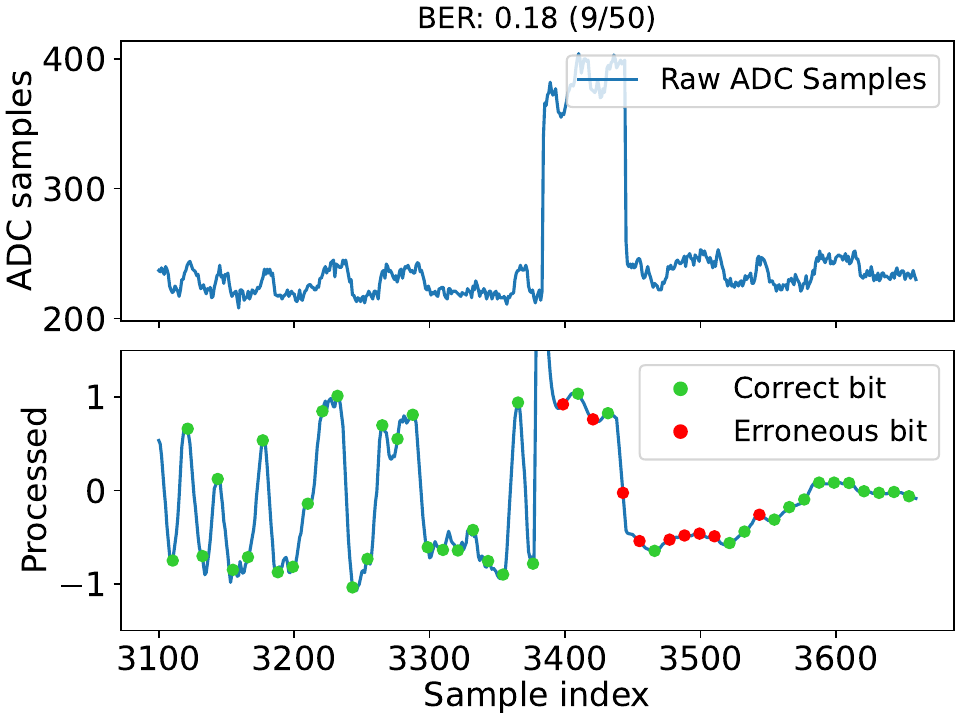}
        \caption{}
        \label{fig:far_transmission_signals_macro_imperfect}
    \end{subfigure}
    \end{minipage}\\
    \caption{Real-world bit transmission experiment towards the Nucleo-G474RE board at \SI{3}{\m}~(top) and \SI{20}{\m}~(bottom) distance from the transmitter. Complete processed receive signals including bit timing ((b)~and~(f)). Raw \ac{ADC} samples and processed signals with bit timing under ideal conditions ((c)~and~(g)) and during self-interference events ((d)~and~(h)).}
    \label{fig:bit_transmission_over_distance}
    \Description{}
\end{figure*}

We now demonstrate that an unmodified, radio-less embedded device can reliably receive digital information over the air -- realizing a functional wireless receiver built entirely from unintended \ac{RF} sensitivities. For this experiment, we configure the Nucleo-G474RE board to use reception path~B and place it at \ac{LOS} distances of \SI{3}{\m} and \SI{20}{\m} from the attacker's antenna and transmit a random binary sequence of approximately \SI{12000}{bits} at a data rate of \SI{1}{kbps} at \SI{868}{\MHz} and a transmit power of \SI{43}{dBm}. At the same time, we record the \ac{ADC} samples from the \ac{MCU}. The setup is shown in~\autoref{fig:bit_transmission_over_distance}~(a) and~(e). The photos are for illustration, as noted earlier, measurements were conducted in an office building.

Using standard software-defined radio processing techniques -- adaptive scaling, DC offset removal, and symbol-timing recovery~\cite{goldsmithWirelessCommunications2005, proakisDigitalCommunications2008} -- we decode the received waveforms, as shown in \autoref{fig:bit_transmission_over_distance}~(b) and~(f). Correctly decoded bits are marked in green, and errors in red. At \SI{3}{\m}, reception is error-free. Even at \SI{20}{\m}, where the received \ac{SNR} is substantially lower, the signal remains clearly distinguishable, with only $781$ bit errors out of $12{,}565$ transmitted bits (\ac{BER} $\approx$ \SI{6.2}{\percent}). We also repeated the experiments in non-\ac{LOS} conditions. With the target device at 3~m distance and a concrete wall in between, the \ac{BER} is $\approx$ \SI{3.1}{\percent}. At 18~m distance with a drywall in between, the BER rises to $\approx$ \SI{15.8}{\percent}.

While the strong transmit signal of \SI{43}{dBm} allows bridging distance and \ac{LOS} obstructions, we emphasize that path loss of around \SI{50}{dB} causes the arrival power to be as small as \SI{-7}{dBm}~($\approx$~\SI{0.2}{\mW}). Crucially, the attacker can reduce transmit power (as evident from \autoref{fig:snr_against_power} and \autoref{fig:ber_vs_tx_power}) in exchange for distance: At 23~dBm, which can be generated using a single-chip RF transceiver~\cite{onsemi_ax5045}, the distance can still be 2~m, while at 17~dBm it would be 1~m. We experimentally validated the latter case and achieved a \ac{BER} of \SI{7.9}{\percent}. We found that the residual bit errors (which could be fully corrected using standard error correction codes~\cite{goldsmithWirelessCommunications2005}) are primarily caused by self-interference from the serial data transmission between the \ac{MCU} and the host computer during sampling, as visible in~\autoref{fig:bit_transmission_over_distance}~(d) and~(h). This interference is an artifact of our measurement setup: in an actual attack scenario, data would be processed locally on the device and such interference would not occur. Further, we observed that other reception paths did not exhibit this sensitivity during data transfer. A deeper investigation of self-interference induced by device activity, as well as techniques to mitigate it, remains subject to future work. Initial results indicate that typical processor workloads, such as ongoing computations, do not interfere with reception.

\section{Countermeasures}
\label{sec:countermeasures}

In this section, we discuss and evaluate potential countermeasures against wireless communication into air-gapped embedded devices, drawing on insights from \ac{EMI} mitigation.

\Paragraph{Shielding}
From a physical perspective, a straightforward defense against \ac{RF} infiltration is electromagnetic shielding, as formalized in TEMPEST standards~\cite{natoTempest2025}, which define strict limits on electromagnetic emissions and mandate containment through shielded enclosures. Because shielding is inherently reciprocal, any barrier that prevents emissions from leaving a device will also block external signals from entering it. However, conventional emission-security tests consider only radiation emitted \emph{by} the device, and therefore do not guarantee protection against signals \emph{entering} the device. 

\begin{figure}
    \centering
    \includegraphics[width=1\linewidth]{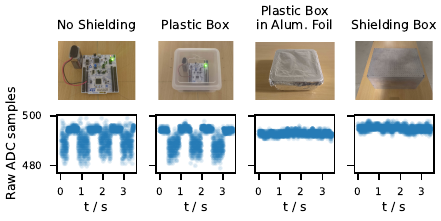}
    \caption{Evaluation of shielding measures.}
    \label{fig:airgap_shielding}
    \Description{}
\end{figure}

To experimentally assess the effectiveness of shielding in mitigating the identified \ac{RF} sensitivities, we repeated the experiment from~\autoref{sec:wiring_impact} with the Nucleo-G474RE board in an air-gapped configuration. The \ac{ADC} recorded samples while the \ac{RF} transmitter was alternatingly switched on and off. We repeated the measurement across four conditions: unshielded, enclosed in a plastic case, the same case wrapped in aluminum foil, and the device placed inside a dedicated shielding box. As shown in~\autoref{fig:airgap_shielding}, introducing a metallic enclosure effectively eliminates the observable reception sensitivity. confirming that shielding can mitigate the phenomenon. While this is one particular example, we emphasize that implementing robust electromagnetic shielding in practice is challenging, demanding systematic testing across reception paths to verify the absence of sensitivities.

\Paragraph{\ac{PCB} Design for Signal Integrity}
To assess how \ac{PCB} design practices can mitigate \ac{RF} sensitivity~\cite{archambeault2002pcb, mohammedIEMIEffectEfficacy2024}, we fabricated two custom two-layer FR4 boards hosting an STM32G474RE \ac{MCU}, on-board power regulation, a USB interface, and multiple \ac{PCB} traces of varying lengths. One board featured a continuous ground plane on the bottom layer, while the other did not, isolating the effect of grounding on \ac{RF} reception.

\begin{figure}
    \centering
    \includegraphics[width=1\linewidth]{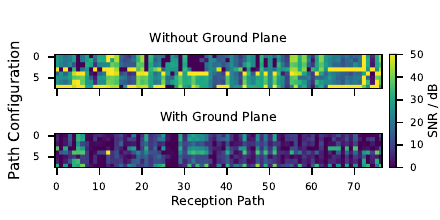}
    \caption{Custom \ac{PCB} sensitivity testing peak \ac{SNR} results for all available reception paths and corresponding \ac{GPIO} configurations.}
    \label{fig:gnd_plane_heatmap}
    \Description{}
\end{figure}

Both boards were evaluated under identical conditions, sweeping all reception paths and the eight preferred \ac{GPIO} configurations. Peak \acp{SNR} are shown in~\autoref{fig:gnd_plane_heatmap}. While both boards exhibit similar structural patterns highlighting the most RF-sensitive configurations, the ground-plane board consistently shows reduced sensitivity. As summarized in~\autoref{tab:case_study_devices}, the ungrounded board exhibited sensitivity on \emph{all} 87 reception paths, compared to 53 paths for the grounded board.

\begin{figure}
    \centering
    \includegraphics[width=1\linewidth]{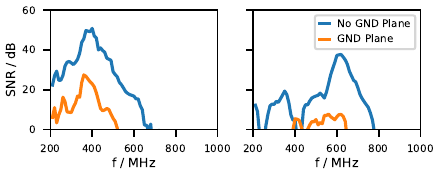}
    \caption{Comparison of reception \ac{SNR} over frequency on the custom \ac{PCB} with and without ground plane for two distinct reception paths.}
    \label{fig:gnd_plane_spectrum}
    \Description{}
\end{figure}

\autoref{fig:gnd_plane_spectrum} compares SNR across frequency for two representative paths. Spectral shapes remain similar, but the ground-plane board achieves more than \SI{20}{\dB} lower sensitivity across the measured range. These results demonstrate that continuous grounding and proper return paths can substantially reduce unintended RF coupling. Nonetheless, residual sensitivities remain under specific configurations, highlighting that grounding alone is insufficient.

\Paragraph{Other Countermeasures}
Additional defenses include monitoring the RF spectrum for anomalous signals, allowing the defender to detect the attack. Another approach is constant jamming of the airgap environment to make it unusable for covert communication. Both measures are complementary to shielding and hardware-level protections. %

\section{Discussion}
\label{sec:discussion}
In the following, we discuss the experimental setup, our results, limitations, and provide directions for future research.

\subsection{Experimental Setup and Results}

In our study, we focused on the necessary steps an attacker needs to perform to receive wireless communication signals: identifying and exploiting \ac{RF} sensitivities in embedded devices. To this end, we used the evaluated devices as data-acquisition platforms, streaming \ac{ADC} samples over USB to a host computer for offline analysis, facilitating controlled experimentation. Consistent with our attacker model, which assumes code execution on the target, we deployed custom firmware on all evaluated platforms. While our proof-of-concept separates signal acquisition and data decoding from normal device operation, a practical air-gap infiltration would integrate the receiver into the device’s legitimate firmware -- a step we consider feasible with moderate additional engineering effort. Unlike exfiltration, which can be performed opportunistically and stealthily, such integration must additionally manage the device's normal operation alongside RF reception, and an imperfect implementation risks observable side effects (\eg, glitches, timing anomalies) that could reveal the compromise. We note that reception can be scheduled selectively, \eg, during idle periods, to minimize such interference and reduce the attack's observability.

For cryptocurrency hardware wallets, which employ software protections to prevent unauthorized firmware modification, we replaced the original \ac{MCU} with an identical but unlocked chip. How attackers gain code execution on deployed systems is orthogonal to our work. In practice, it could result from established attack vectors such as boot-chain attacks~\cite{cuiBADFETDefeatingModern2017, wouters2022blackhat} or supply-chain compromise, as previously discussed for hardware wallets~\cite{roth2018ccc}. Our study shows that unintended \ac{RF} sensitivities are a common physical property of embedded devices, including security-critical platforms. These findings should therefore not be interpreted as a comparative security evaluation of individual products or vendors. To isolate the underlying physical mechanisms, all measurements were performed on bare \acp{PCB} and evaluating the effect of enclosures, shielding, and attached peripherals is left to future work.

For sensitivity assessment, we assume a priori knowledge of the radio signal, \ie, we recorded labeled observations with the \ac{RF} signal turned on and off. This differs from the communication scenario, in which an attacker must detect and decode signals without prior knowledge. However, this controlled setup is necessary to systematically identify and characterize \ac{RF} sensitivities in a laboratory setting. As our end-to-end receiver demonstration shows, the effects are sufficiently strong for blind signal detection and reliable wireless communication at low error rates.

As in any wireless system, the received signal strength must exceed the receiver’s minimum sensitivity. Our results show that, for an ordinary embedded device, reliable reception only requires an arrival power as low as \SI{-7}{dBm}, relying solely on unintended board- and chip-level coupling. Despite such minimalistic capabilities, the link budget is sufficient for long-distance communication. The attacker can trade power for distance and is not fundamentally constrained in transmit power.

\subsection{Limitations}

\textbf{ADC Availability.}
A fundamental precondition for exploiting RF sensitivities on radio-less embedded devices is the presence of a coupling interface between the analog and digital domains, \ie, an \ac{ADC} that the attacker has access to. Without an ADC, the attacker cannot obtain measurable digital values corresponding to analog \ac{RF} signal variations. 

\textbf{Receiver Functionality.}
As shown, radio-less embedded devices can receive digital communication signals over substantial distances (tens of meters). This requires ($i$)~that sufficient \ac{RF} energy arrives at the target and ($ii$)~that the compromised device exhibits sensitivity in the corresponding frequency range. The attack is therefore limited to scenarios where both conditions are satisfied. While some improvement in reception sensitivity may be possible, we expect potential gains to be limited due to the unconventional approach to signal reception. 
Further, unlike standard radio receivers, the functionality leveraged here is not frequency-selective, \ie, there is no local oscillator on no band selection. Frequency selectivity is determined solely by the frequency response of the specific reception path, making the receiver potentially susceptible against strong near-by radio interference. %

\textbf{Black-Box Scenarios.} In this work, we have demonstrated practicality of receiving radio communication signals via unintended electromagnetic sensitivities of embedded devices. Our analysis assumes that the attacker possesses enough knowledge about target devices to determine their sensitivities. In black-box scenarios where a reference device is unavailable -- such as when targeting custom hardware platforms -- the attacker cannot identify optimal transmission frequencies and device configurations. Then, the attacker could randomly iterate over reception paths on the target device, while transmitting broadband signals that could cover multiple sensitive frequencies in parallel. If real-time feedback, \eg, via \ac{RF} exfiltration, from the target device is available, the attacker can further refine both the receiver configuration and transmission parameters through in-situ optimization, adapting based on whether signals are successfully received.

\Paragraph{Co-Existence of Functionalities}
We reconfigure the peripheral settings of the \ac{MCU} to connect sensitive traces to an \ac{ADC}, enabling radio reception. This disables the pin’s native functionality, such as bus communication, during reception. A promising direction for future work is to preserve pin functionality while still enabling \ac{RF} reception. This would require a mechanism to simultaneously route the signal to both the peripheral device, \eg, \ac{I2C} or \ac{SPI}, and the \ac{ADC}. An attacker could then attempt to isolate radio-induced effects from the raw bus signals, potentially allowing continuous or passive reception without disrupting normal operation.

\subsection{Future Work}

In this work, we showed that subtle \ac{RF} sensitivities of standard embedded devices can be exploited to realize covert radio receivers, \eg, for command-and-control of compromised air-gapped systems. Our initial exploration of the mechanism opens up several promising directions for future work.

\Paragraph{Transmitted Waveforms and Modulation Schemes}
In our experiments, we validated receiver sensitivity using single-carrier \ac{RF} signals and demonstrated successful reception of \ac{OOK}-modulated data. Further work is needed to assess the feasibility of using more complex waveforms, such as pulsed \ac{UWB}, frequency or phase-modulated, and multi-carrier signals.

\Paragraph{Other Platforms}
In our study, we focused on reception using STM32 \acp{MCU}, a microcontroller family found across a broad spectrum of commercial devices. Porting the mechanism to other MCU families is left for future work. Further, \acp{FPGA} constitute an interesting  target for sensitivity assessment, given their large configuration space. We also believe that the approach could extend to other platforms such as smartphones and laptops. However, further work is required to assess whether data exposed at the operating-system level offers sufficient fidelity for reliable signal reception.

\Paragraph{Bidirectional Communication}
We leave the demonstration of bidirectional wireless communication between two air-gapped devices for future work. From a physical standpoint, such a setup is clearly viable -- for instance, prior work like NoiseSDR~\cite{camuratiNoiseSDRArbitraryModulation2022} demonstrates how signal transmission can be realized using standard embedded devices. Integrating such a transmitter with our reception mechanism poses no conceptual barrier, requiring additional system integration and tuning.

\Paragraph{High EM-activity environments}
Our experiments took place in environments with ambient wireless activity, but not in high-activity EM environments such as data centers, where the target receiver would operate under significant interference. While such interference can be modeled statistically, experimental validation is left for future work.

\section{Related Work}
\label{ref:relatedwork}

In this section, we give an overview of relevant literature on \ac{IEMI} applications in a security context and airgap research. Moreover, we summarize prior works on turning embedded devices into radio receivers and elaborate how our work differs from these proposals.

\Paragraph{\ac{IEMI} and Sensor Security} Most research on \ac{IEMI} has focused on malicious remote manipulation of sensors, showing that a wide range of modalities can be affected, including temperature sensors~\cite{tuTrickHeatManipulating2019}, accelerometers and gyroscopes~\cite{pahlIntendedElectromagneticInterference2021}, cameras~\cite{renGhostShotManipulatingImage2025,jiangGlitchHikerUncoveringVulnerabilities2023,kohlerSignalInjectionAttacks2022}, voltage and current sensors~\cite{yangReThinkRevealThreat2025}, capacitive touchscreens~\cite{wangGhostTouchTargetedAttacks2022}, speed sensors~\cite{shoukryNoninvasiveSpoofingAttacks2013}, barometric pressure sensors~\cite{lavauImpactIEMIPulses2022,lavauSusceptibilitySensorsIEMI2021}, analog sensors in biomedical implants~\cite{kuneGhostTalkMitigating2013}, and infrared sensors~\cite{selvarajElectromagneticInductionAttacks2018}. A related line of research investigates \ac{IEMI} to tamper with device-internal communication buses~\cite{jangParalyzingDronesEMI2023, zhangElectromagneticSignalInjection2023}. Adversaries have also leveraged infrastructure and peripherals, using fluorescent lamps~\cite{yangLightAntennaCharacterizingLimits2025}, wired headphones~\cite{kasmiIEMIThreatsInformation2015}, or grounding paths~\cite{jiangPowerRadioManipulateSensor2025} to aid \ac{IEMI}-based sensor manipulation. While not related to \ac{IEMI}, prior research has demonstrated that sensors can be exploited for covert data transfer, \eg, using microphones, light, and magnetic sensors on smartphones to transfer digital data~\cite{hasanSensingenabledChannelsHardtodetect2013,subramanianExaminingCharacteristicsImplications2013}.

\Paragraph{Bridging the Airgap} As discussed in the Introduction, airgap infiltration has received relatively little attention, with prior work by Guri~\etal~\cite{guriBitWhisperCovertSignaling2015, guriMosquitoCovertUltrasonic2018, guriAIRJumperCovertAirgap2019},  Kuhnapfel~\etal~\cite{kuhnapfelLaserSharkEstablishingFast2021}, and Kasmi~\etal~\cite{kasmiAirgapLimitationsBypass2016}. 

Exfiltration of data from air-gapped system has first been studied by Kuhn and Anderson~\cite{kuhnSoftTempestHidden1998}. Their \enquote{Soft TEMPEST} approach used a display to control the amplitude of emitted \ac{EM} signals from software. Since then, many subsequent works, \eg, \cite{guriAirHopperBridgingAirgap2014, guriBitWhisperCovertSignaling2015, guriGSMemDataExfiltration2015, shenWhenLoRaMeets2021, guriDiskfiltrationAcousticData2016, agadakosJumpingAirGap2017, guriAirGapElectromagneticCovert2024, zhanBitJabberWorldsFastest2020, carraraAcousticCovertChannels2015}, have studied various hardware components and leakage sources. In the context of electromagnetic leakage, Camurati~\etal~\cite{camuratiNoiseSDRArbitraryModulation2022} presented Noise-SDR, allowing use complex modulation schemes on device-internal leakage sources. In contrast, SpiralSpy~\cite{liSpiralSpyExploringStealthy2022} and DiskSpy~\cite{xuDiskSpyExploringLongRange2025} are passive approaches that rely on cooling fans and hard disk vibrations, respecitvely, to modulate externally applied radar signals.

\Paragraph{Embedded Devices as Radio Receivers and Differentiation from Previous Work} Kasmi~\etal~\cite{kasmiAirgapLimitationsBypass2016} and Esteves~\cite{lopesestevesIntroductionIntentionalElectromagnetic2025,estevesElectromagneticInterferenceInformation2023} first explored the use of \ac{IEMI} for covert communication, employing kilowatt-range radio transmissions to manipulate temperature sensor readings at data rates of roughly 2.5~bit/s. To the best of our knowledge, this remains the only work to leverage \ac{IEMI} for wireless communication across the airgap. In contrast, our reception mechanism does not rely on the availability of sensors and furthermore exhibits sensitivity to \ac{RF} signals with an arrival power as low as \SI{0.2}{\mW}, reducing the required transmit power by several orders of magnitude compared to prior work while achieving data rates in the kilobit range.

Several projects have demonstrated creative ways of realizing \acp{SDR} by augmenting embedded platforms with external \ac{RF} front-end circuitry such as antennas, filters, and amplifiers. Examples include ARM Radio on STM32 \acp{MCU}~\cite{diBeneArmRadio,garlassiArmRadio2020}, PiccoloSDR~\cite{aufrancPiccoloSDR2021} and PicoSDR~\cite{dvorakPicoSDR2025} on the Raspberry Pi Pico, and FPGA-based receivers~\cite{dawsonjonFPGARadio,newhouseBLEFPGA}. These works showcase the flexibility of embedded hardware, but share common requirements: external components (at least antennas, often amplifiers or filters), high-speed sampling and digital signal processing, and high resource usage to realize radio reception. Our approach is fundamentally different: we show that completely unmodified, sensor-less embedded devices can act as covert radio receivers with no external hardware and minimal signal processing requirements for realistic attack scenarios on low-resource embedded platforms.

\section{Conclusion}
\label{sec:conclusion}

In this paper, we studied the potential of parasitic \ac{RF} sensitivities in conventional embedded systems to receive wireless communication signals. We introduced a systematic methodology for discovering such sensitivities and evaluated it across 14 devices, finding exploitable reception behavior in all of them. Our results show that an adversary with code execution on an air-gapped embedded device can leverage these unintended sensitivities to receive modulated \ac{RF} signals, enabling command-and-control over tens of meters and even under non-line-of-sight conditions. These findings demonstrate that the absence of native radio capabilities does not imply physical isolation, and they motivate a reassessment of air-gap security assumptions as well as the design of embedded systems with improved resilience against \ac{EMI}.

\section*{Acknowledgments}
We thank Ilyes Lajili for his assistance with data collection. This work was in part funded by the Deutsche Forschungsgemeinschaft (DFG, German Research Foundation) under Germany’s Excellence Strategy - EXC 2092 CaSa - 390781972.

\bibliographystyle{ACM-Reference-Format}
\balance
\bibliography{airgap_rx_ccs}

\appendix %

\section{Open Science} %

We release the following artifacts to assist replication of our findings: ($i$)~\ac{MCU} firmwares for acquiring \ac{ADC} samples and streaming to the host computer, ($ii$)~Python code for device control, reading samples, and conducting sensitivity testing, ($iii$)~design files and manufacturing information for our custom \acp{PCB}, and ($iv$)~software-defined receiver implementation and data to demodulate signals received by the Nucleo-G474RE board and calculate \ac{BER}. The artifact is available at \url{https://doi.org/10.5281/zenodo.18340914}.

\section{Ethical Considerations}
Our research explores offensive capabilities with potential indirect impacts on individuals, organizations, and society.

\textbf{Stakeholders and Potential Impacts.}
\textit{Device manufacturers} face disclosure of parasitic electromagnetic sensitivities arising from physical properties rather than architectural flaws, informing future design and mitigation. \textit{End users} benefit from increased awareness and improved electromagnetic resilience, as communicating with an already-compromised system poses no greater risk than the initial compromise. \textit{Researchers, standards organizations, and society} gain shared methodologies and heightened awareness of air-gap threats. \textit{Malicious actors} could gain command-and-control over compromised air-gapped devices, contingent on prior compromise and unauthorized firmware modification. \textit{Nearby individuals} could be affected by emissions causing service disruptions.

\textbf{Mitigations.}
All experiments used author-owned devices and purpose-built firmware. We did not circumvent firmware protections, interfere with live systems, or process personal data. Released firmware implements no product behavior and cannot be directly weaponized, secure boot prevents unauthorized firmware loading. Experiments were conducted in closed environments without uninvolved individuals. We evaluated countermeasures and followed responsible disclosure with affected manufacturers.

\textbf{Decision to Proceed.}
Benefits outweigh harms: our work illuminates a previously unrecognized attack vector, motivates mitigations, and supports proactive electromagnetic security evaluation, while immediate user risk remains minimal due to firmware validation mechanisms.

\section{Recommended Path Configurations}
\label{sec:recommended_path_configurations}

See the recommended path configurations in~\autoref{tab:recommended_path_configurations}.

\begin{table}[h]
\caption{Recommend path configurations.}
\label{tab:recommended_path_configurations}
\centering
{\fontsize{6}{7}\selectfont\def\arraystretch{2}\tabcolsep=3pt
\begin{tabular}{llllll}
    \toprule
    & \textbf{PUPD} & \textbf{Output Value} & \textbf{Mode} & \textbf{Output Type} & \textbf{\makecell[l]{Output Speed\\(Not Relevant)}} \\
    \midrule
    1 & Pull-Down & High & Input & Open-Drain & Medium \\ 
    2 & Pull-Down & High & Output & Open-Drain & Medium \\ 
    3 & Pull-Down & High & Alternate Function & Open-Drain & Medium \\ 
    4 & Pull-Down & High & Analog & Open-Drain & Medium \\ 
    5 & Pull-Up & Low & Input & Open-Drain & Medium \\ 
    6 & Pull-Up & Low & Output & Open-Drain & Medium \\ 
    7 & Pull-Up & Low & Alternate Function & Open-Drain & Medium \\ 
    8 & Pull-Up & Low & Analog & Open-Drain & Medium \\
    \bottomrule
\end{tabular}
}
\end{table}

\section{USB vs. Battery-Powered Operation}
\begin{figure}[htb]
    \centering
    \includegraphics[width=0.7\linewidth]{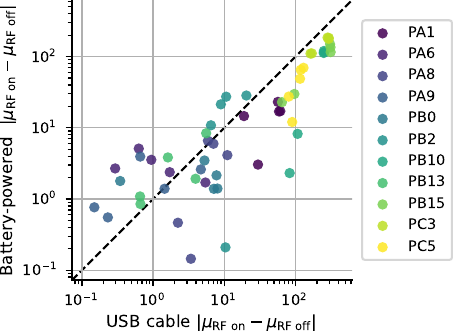}
    \caption{Mean \ac{ADC} difference between RF on and RF off under USB power versus battery power, for 11 reception paths (GPIO pins) and five GPIO configurations at a single frequency. Points near the diagonal indicate equivalent sensitivity.}
    \label{fig:usb_vs_battery}
    \Description{}
\end{figure}
\autoref{fig:usb_vs_battery} shows that sensitivities measured with the USB cable attached closely match those measured under battery power, indicating that cabled testing is representative of air-gapped conditions.

\section{Inter-Device Behavior}
\label{sec:inter_device_qty}

\begin{figure}[htb]
    \centering
    \includegraphics[width=1\linewidth]{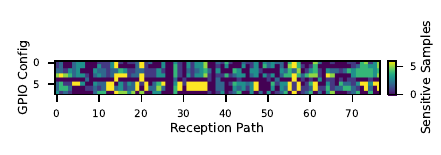}
    \caption{Number of sensitive device samples as a function reception paths and paths configuration.}
    \label{fig:inter_device_qty}
    \Description{}
\end{figure}

\autoref{sec:inter_device} describes the experiment to assess whether reception sensitivities persist across six identical samples of Nucleo-G474RE boards. For each, we run a path-configuration sweep across frequency and use a \ac{SNR} threshold of \SI{12}{\dB} to decide a sensitivity. \autoref{fig:inter_device_qty} shows the number of sensitive samples over reception paths and configurations. Here, we can see that sensitivities are repeatable across all samples for various reception paths, especially for GPIO configurations 5, 6~and 7, \cf~\autoref{tab:recommended_path_configurations}.

\section{Common Ground Connection}
\label{sec:common_gnd}
\begin{figure}[htb]
    \centering
    \includegraphics[width=1\linewidth]{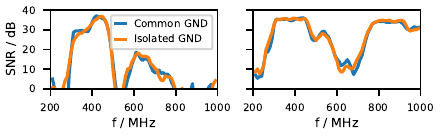}
    \caption{With and without common ground between transmitter and receiver.}
    \label{fig:conductive}
    \Description{}
\end{figure}

\autoref{fig:conductive} shows the \ac{SNR} of two reception paths across frequency with and without a common ground between the \ac{RF} transmitter and the \ac{DUT}. In the common-ground configuration, both devices share the same computer for control and power (\autoref{fig:exp_setup}). In the isolated case, the \ac{DUT} is powered by a separate battery-powered laptop for galvanic isolation, with all other conditions identical. The sensitivity profiles are virtually unchanged across both configurations. The observed sensitivities therefore do not stem from unintended conductive coupling, but from genuine interaction between the \ac{DUT} and propagated electromagnetic signals.

\section{Effect of ADC Oversampling}
\label{sec:appendix_adc}

\begin{figure}[htb]
    \centering
    \includegraphics[width=1\linewidth]{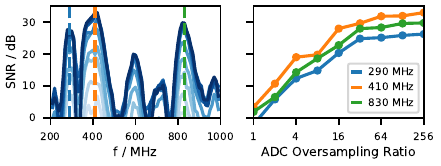}
    \caption{Effect of the \ac{ADC} oversampling ratio on reception sensitivity \ac{SNR}.}
    \label{fig:oversampling}
    \Description{}
\end{figure}

\ac{ADC} oversampling increases the number of samples acquired per signal period, which can improve the effective \ac{SNR}. As shown in~\autoref{fig:oversampling}, the sensitive frequency regions remain largely unchanged across all oversampling ratios (left), indicating that oversampling does not shift the device's inherent spectral response. The \ac{SNR} in~dB instead improves almost linearly with every doubling of the ratio (right), plateauing at a factor of approximately~32. Oversampling can therefore improve received signal quality, allowing an attacker to trade transmission rate against reception reliability.

\section{Device Orientations}
\label{sec:device_orientations}

\begin{figure}[htb]
    \centering
    \includegraphics[width=1\linewidth]{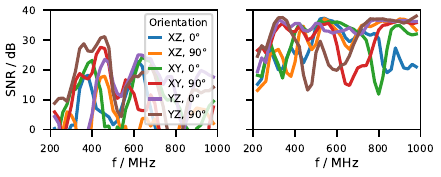}
    \caption{\ac{SNR} over frequency for two reception paths with the device oriented in the $xz$, $xy$, and $yz$ planes, each rotated by \SI{0}{\degree} and \SI{90}{\degree}.}
    \label{fig:snr_vs_planes}
    \Description{}
\end{figure}

We placed the device in the $xz$, $xy$, and $yz$ planes and rotated it by \SI{0}{\degree} and \SI{90}{\degree}. The resulting frequency responses are shown in~\autoref{fig:snr_vs_planes}. While the detailed spectral characteristics vary across orientations, the sensitivities remain significant and correlated in frequency, indicating that orientation affects but does not eliminate the phenomenon.

\end{document}